\documentclass[twocolumn,a4paper,reprint,amssymb,aps,prd,nofootinbib,groupedaddress,superscriptaddress]{revtex4-1}

\makeatletter
\renewcommand{\p@subsection}{}
\renewcommand{\p@subsubsection}{}
\makeatother

\usepackage{soul}
\usepackage{graphicx}
\usepackage{dcolumn}
\usepackage{bm}
\usepackage[colorlinks=true, linkcolor=blue, citecolor=red]{hyperref}
\usepackage{floatrow}
\newfloatcommand{capbtabbox}{table}[][\FBwidth]

\usepackage{blindtext}
\usepackage[normalem]{ulem}
\usepackage{graphicx}  % needed for figures
\usepackage{dcolumn}   % needed for some tables}
\usepackage{bm}        % for math
\usepackage{amsmath,amssymb}
\usepackage{hyperref}
\usepackage{color}
\definecolor{purple}{rgb}{0.58,0.0,0.83}
\usepackage[toc,page]{appendix}

\begin{document}

\title{Cosmological parameter inference with Bayesian statistics}
\author{Luis E. Padilla}  
%\email{epadilla@fis.cinvestav.mx}
\affiliation{Departamento de F\'isica, Centro de Investigaci\'on y de Estudios Avanzados del IPN, A.P. 14-740, 07000 M\'exico D.F.,
  M\'exico.}
  \affiliation{Department of Astronomy and Texas Cosmology Center, University of Texas, Austin, TX, 78712-1083, U.S.A.}
\author{Luis O. Tellez}  
\affiliation{Departamento de F\'isica, Centro de Investigaci\'on y de Estudios Avanzados del IPN, A.P. 14-740, 07000 M\'exico D.F.,
  M\'exico.}
  \affiliation{Instituto de Ciencias F\'isicas, Universidad Nacional Aut\'onoma de M\'exico, 
Apdo. Postal 48-3, 62251 Cuernavaca, Morelos, M\'exico.}
    \author{Luis A. Escamilla}  
\affiliation{Departamento de F\'isica, Centro de Investigaci\'on y de Estudios Avanzados del IPN, A.P. 14-740, 07000 M\'exico D.F.,
  M\'exico.}
  \affiliation{Instituto de Ciencias F\'isicas, Universidad Nacional Aut\'onoma de M\'exico, 
Apdo. Postal 48-3, 62251 Cuernavaca, Morelos, M\'exico.}
 \author{J. Alberto Vazquez}  
\email{javazquez@icf.unam.mx}
\affiliation{Instituto de Ciencias F\'isicas, Universidad Nacional Aut\'onoma de M\'exico, 
Apdo. Postal 48-3, 62251 Cuernavaca, Morelos, M\'exico.}
\affiliation{Departamento de F\'isica, Centro de Investigaci\'on y de Estudios Avanzados del IPN, A.P. 14-740, 07000 M\'exico D.F.,
  M\'exico.}
\date{\today}
\newcommand{\lp}[1]{\textcolor{purple}{(lp: #1)}}

\begin{abstract}

Bayesian statistics and Markov Chain Monte Carlo (MCMC) algorithms have found 
their place in the field of Cosmology. They have become important mathematical 
and numerical tools, especially in parameter estimation and model comparison. 
In this paper, we review some fundamental concepts to 
understand Bayesian statistics and then introduce MCMC algorithms 
and samplers that allow us to perform the parameter inference procedure.
We also introduce a general description of the
standard cosmological model, known as the $\Lambda$CDM model,
along with several alternatives, and current datasets coming 
from astrophysical and cosmological observations. 
Finally, with the tools acquired, we use an MCMC algorithm implemented in python
to test several cosmological models and find out the combination of parameters
that best describes the Universe.

\end{abstract}

\maketitle
%%%================================================================%%%
\section{Introduction}
%%%================================================================%%%

The beginning of the standard cosmology as it is known today emerged after 1920, when 
the Shapley-Curtis debate was carried out \cite{debate}. This debate was held between 
the astronomers Harlow Shapley and Heber Curtis, resulting in a revolution 
for astronomy at that time by reaching an important conclusion: 
``The Universe had a larger scale than the Milky Way''. Several observations at that epoch
established that the size and dynamics of the cosmos could be explained by Einstein's General 
Theory of Relativity. 
In its childhood, cosmology was a speculative science based only on a few datasets, 
and it was characterized by a dispute between two cosmological models: the steady state model 
and the Big Bang (BB) theory. 
It was not until 1990, when the amount of data increased enough to discriminate and 
rule out compelling theories, that the BB model awarded was the most accepted. 
During the same decade, David Schramm heralded the ``Golden Age of Cosmology'' at a National 
Academy of Sciences colloquium \cite{goldenage}. 

Once the new age of cosmological observations arrived with a large variety of data, it was 
necessary to confront the cosmological models with such data, usually done through statistics. 
It is important to stress out that, since we have a unique Universe, we cannot rely on a 
frequentist interpretation of statistics (we are not able to create multiple 
Universes and make a frequentist inference of our models). 
An alternative approach that will help in our task is the Bayesian statistics. 
In Bayesian statistics, the probability is interpreted as a ``degree 
of belief'', and it may be useful when repetitive processes are complicated to~reproduce. 

The main aim of this work is to provide an introduction of Bayesian parameter inference 
and its applications to cosmology. We assume the reader is familiarized with the basic 
concepts of statistics, but not necessarily with Bayesian statistics. Then, we provide 
a general introduction to this subject, enough to work out some examples. 
This review is written in a generic way so that the parameter inference 
theory may be applicable to any subject, in particular we put into practice the Bayesian 
concepts into the field of cosmology.

The paper is organized as follows. In Section \ref{sec:Bayesian}, we point out the
main differences between the Bayesian and Frequentist approaches. Then, in Section 
\ref{sec:first}, we introduce the basic mathematical concepts in Bayesian statistics to perform 
the parameter estimation procedure for a given model. Once we have the mathematical 
background, we continue, in
Section \ref{sec:tools}, with some numerical resources that are able to simplify our task,
 especially for models with several 
parameters that need to be tested with many datasets. 
With these methods and tools in place, we provide the example of fitting a straight line in Section \ref{sec:line}.
In Section \ref{sec:Cosmology}, we present an introduction to cosmology, 
and then, in Section \ref{example1}, 
focus on some codes to compute the cosmological observables. 
In Section \ref{sec:Simplemc}, we constrain the 
parameter space that describes 
the standard cosmological model, namely the $\Lambda$CDM model, along with several extensions. 
Finally, in Section \ref{sec:Conclusions}, we present our conclusions.

%%%================================================================%%%
\section{Bayesian vs Frequentist statistics}
\label{sec:Bayesian}
%%%================================================================%%%

Fundamentally, the main difference between Bayesian and Frequentist statistics is 
on the definition of probability. From a Frequentist point of view, 
probability has meaning in limiting cases of repeated measurements
\begin{equation}
    P=\frac{n}{N},
\end{equation}
where $n$ denotes the number of successes, and $N$ the total number of trials. 
Frequentist statistics defines the probability $P$ as the limit of the number of independent 
trials going to infinity. 
Then, {{for Frequentist} statistics, probabilities are fundamentally related to 
frequencies of events}. On the other hand, in Bayesian statistics, the concept of 
probability is extended to cover degrees of certainty about a statement. 
{{For Bayesian statistics}, probabilities are fundamentally related to our 
knowledge concerning an event}.
%MDPI: Please confirm whether the bold are necessary.
% Authors: They're not necessary, but we'd like to stress out these important concepts.

Here, we introduce some key concepts to understand the consequences this 
difference entails; for an extended review see Reference \cite{bayeslecture, AlanH, RobT, LiV, RobTr, Jaffe:1995qu, Verde:2003ey, DAgostini:1995jqe, 2017ARA&A..55..213S, Liddle:2004nh, parametros} and references therein. 
\noindent
Let $x$ be a random variable related to a particular event and $P(x)$ its corresponding
probability distribution, for both cases, the same rules of {probabilities 
apply} \footnote{These rules are defined for a continuous variable; however, the 
corresponding discrete definition can be given immediately by replacing 
$\int dx \rightarrow \sum$.}:
%MDPI: Subequation is not allowed, please confirm and modify.
%Authors: Subequations were removed.

\begin{equation}\label{rule1}
    P(x)\geq 0,
\end{equation}
\begin{equation}\label{rule2}
\int_{-\infty}^\infty dxP(x)=1.
\end{equation}

For {mutually exclusive} events, we have
%MDPI: Please confirm whether the italics are necessary. 
%Authors: For 'mutually exclusive', they are not necessary.
%

\begin{equation}\label{rule3}
    P(x_1\cup x_2)=P(x_1)+P(x_2), 
\end{equation}
but, in general

\begin{equation*}
    P(x_1\cup x_2)=P(x_1)+P(x_2) - P(x_1 \cap x_2).
\end{equation*}

These rules are summed up as follows:
the first condition (\ref{rule1}) is necessary due to the probability of having an
event is always positive; the second rule (\ref{rule2}) is a normalized relation, 
which tells us that we are certain to obtain one of the possible outcomes;
now, in the third point (\ref{rule3}) we have that the probability of obtaining an observation,
from a set of mutually exclusive events, is given by the individual probabilities of each event;
finally, and in general, if one event occurs given the occurrence of another
then the probability that both $x_1$ and $x_2$ happen is equal to the 
probability of $x_1$ times the probability of $x_2$ given that $x_1$ has already happened 
\begin{equation}\label{rule4}
    P(x_1\cap x_2)=P(x_1)P(x_2|x_1).
\end{equation}

If two events $x_1$ and $x_2$ are mutually exclusive, then 
\begin{equation}
    P(x_1\cap x_2) = 0 = P(x_2\cap x_1).
\end{equation}

The rules of probability distributions must be fulfilled by both Frequentist and Bayesian
statistics. However, there are some consequences derived by the fact that these two scenarios 
have a different definition of probability, as we shall see below.

%%%------------------------------------------------------------------------%%%
\subsection{Frequentist statistics} \label{freq}
%%%------------------------------------------------------------------------%%%

{Any frequentist inferential procedure relies on three basic ingredients: the data, 
the model and an estimation procedure. 
The main assumption in Frequentist statistics is that the data has 
a definite, albeit unknown, underlying distribution to which all inference~pertains.}

The \textbf{{data}} is a measurement or observation, denoted by $X$, that can take 
any value from a corresponding sample space. A\textbf{ {sample space} }of an 
observation $X$ can be defined as a measurable space $(x,\hat B)$ that contains 
all values that $X$ can take upon measurement.
In Frequentist statistics, it is considered that there is a probability function 
$P_0:\hat B\rightarrow [0,1]$ in the sample space $(x,\hat B)$ representing the ``true 
distribution of the data'' \[X\sim P_0.\]

For Frequentist statistics, the \textbf{{model}} $Q$ is a collection of
probability measurements $P_\theta:\hat B\rightarrow[0,1]$ in the sample space $(x,\hat B)$. 
The distributions $P_\theta$ are called \textit{model distributions}, with $\theta$ being the model parameters; in this approach $\theta$ is unchanged. 
A model $Q$ is said to be well-specified if it contains the true distribution of the 
data $P_0$, i.e., \[P_0\in Q.\]

Finally, we need a point-estimator (or estimator) for $P_0$. An \textbf{{estimator}} 
for $P_0$ is a map $\hat P:x\rightarrow Q$, representing our best guess 
$\hat P\in Q$ for $P_0$ based on the data $X$.
Therefore, the Frequentist statistics is based on trying to answer the following questions: 
``what the data is trying to tell about $P_0$?'' or ``considering the data, 
what can we say about the mean value of $P_0$?''

%%%------------------------------------------------------------------------%%%
\subsection{Bayesian statistics}
%%%------------------------------------------------------------------------%%%

In Bayesian statistics, data and model are two elements 
of the same space \cite{bayeslecture}, 
i.e., no formal distinction is made between measured quantities $X$ and 
parameters $\theta$. One may envisage the process of generating a measurement's outcome $Y=y$ as 
two draws, one draw for $\Theta$ (where $\Theta$ is a model with associated probabilities 
to the parameter $\theta$) to select a value of $\theta$ and a subsequent 
draw for $P_\theta$ to arrive at $X=x$. This perspective may seem rather absurd when thinking in a Frequentist way, but, in Bayesian statistics, where
probabilities are related to our own knowledge, it results natural to associate probability
distributions to our parameters. In this way, an element $P_\theta$ of the model is interpreted 
simply as the distribution of $X$ given the parameter value $\theta$, i.e., as the conditional
distribution $X|\theta$.

%%%------------------------------------------------------------------------%%%
\subsection{Comparing both descriptions}
%%%------------------------------------------------------------------------%%%

\begin{table}[t!]
%\centering
\begin{center}
\resizebox{9cm}{!} {
\begin{tabular}{l|l} 
 \hline
\qquad \textbf{Frequentist} &\qquad  \textbf{Bayesian} \\ 
\cline{1-2}\noalign{\smallskip}
 
 \qquad Data are a repeatable random  \qquad & \qquad Data are observed from  \qquad \\ 
 \qquad sample. There is a frequency. \qquad & \qquad the realized sample.              \qquad \\
 \noalign{\smallskip}
 \hline 
 \noalign{\smallskip}
\qquad  Underlying parameters remain \qquad & \qquad Parameters are unknown and   \qquad \\
\qquad  constant during this repeatable \qquad & \qquad described probabilistically. \qquad \\
\qquad process. &  \\
\noalign{\smallskip}
\hline
\noalign{\smallskip}
\qquad Parameters are fixed. \qquad & \qquad Data are fixed. \\ [1ex] 
 \hline
 \hline
\end{tabular}
\caption{\footnotesize{Main differences between the Bayesian and Frequentist interpretations.}}
\label{table:description}
}
\end{center}
\end{table}

Table \ref{table:description} provides a short summary of the most important differences 
between the two statistics.
To understand these differences, let us review a basic example.
Here, we present an experiment and, since we are interested in comparing both
descriptions, we show only the basic results from both points of view: 
Frequentist and Bayesian. 

\textit{Example.-} {Let us assume} we have a coin that has a probability $p$ to land as 
heads and a probability $1-p$ to land as tails. Our goal is to know whether this coin is fair ($p=0.5$) or not. In the process of trying to estimate $p$, we flip the 
coin 14 times, obtaining heads in 10 of the trials. 
Now, we are interested in the next two possible events. To be precise: 
``What is the probability that in the next two tosses we will get two heads in a row?''
\begin{itemize}
\item \textit{Frequentist approach}. As mentioned previously, in Frequentist 
statistics probability is related to the frequency of events, then our best estimate 
for $p$ is $P(head)=p=\frac{\# \: of\: heads}{\# \: of\: events}=10/14$. 
So, the probability of having 2 heads in a row is $P(2heads)=P(head)P(head)\simeq 0.51$. 
\item \textit{Bayesian approach}. In Bayesian statistics, $p$ is not a value; it is a random 
variable with its own distribution, and it must be defined by the existing evidence 
(the 14 trials and 10 successes). 
Then, by considering that we do not know anything about $p$ a priori and 
averaging over all possible values of $p$, we have that the probability of having two heads is
\begin{equation}
P(2heads|D)=0.485.
\end{equation}
This Bayesian example will be expanded in detail during the following section,
but, for now, we just want to stress out that both approximations arrive at different results.
\end{itemize}

In the Frequentist approach, since we adopt the probability as a frequency of events 
(the probability of having a head was fixed by $p=10/14$); hence, the final result 
was obtained by only multiplying each of these probabilities (since we assume the 
events are independent of each other). On the other hand, in the Bayesian 
framework, it was necessary to average over all possible values of $p$ in order to 
obtain a numerical value. However, in both cases, the probability differs from the 
real one ($P(2heads)=0.25$) because we do not have enough data for our estimations.

In this example, we have seen the forward application of statistics: using a mathematical model to relate measured quantities to an unknown quantity of interest, in this case, the probability of getting two heads in a row. This can also be applicable to the inverse problem. That is, by the having the data, we would like to obtain information about the parameters of a given model \cite{2002AIPC..617..477M, AlanH}. In Section \ref{models}, we will illustrate this point by using different cosmological observations and finding out the best-fit values of the parameters that describe a given model.

%%%================================================================%%%
\section{A first look at Bayesian statistics}
\label{sec:first}
%%%================================================================%%%

Before we start with the applications of Bayesian statistics in cosmology, it is advisable 
to understand the important mathematical tools within the Bayesian procedure. 
In this section, we present a basic revision but encourage the reader to
look for the formal treatment in the literature, cited in each section.   

%%%------------------------------------------------------------------------%%%
\subsection{Bayes' theorem, priors, posteriors and all that stuff}
\label{sub:BTPP}
%%%------------------------------------------------------------------------%%%

When anyone is interested in the Bayesian framework, there are several concepts 
to understand before presenting the results. In this section, we 
review these concepts, and then we get back to the example of the coin toss given 
in the last section. 

\textbf{{The Bayes' theorem.}} The Bayes' theorem is a direct consequence of the 
axioms of probability shown in Equations \eqref{rule1} - \eqref{rule4}. From Equation \eqref{rule4}, 
without loss of generality, it must be fulfilled that $P(x_1\cap x_2)=P(x_2\cap x_1)$. 
In such a case, the following relation applies:
\begin{equation}
P(x_2|x_1)=\frac{P(x_1|x_2)P(x_2)}{P(x_1)}.
\end{equation}

As already mentioned, in the Bayesian framework, data and model are part of the same space. 
Given a model (or hypothesis) $H$, considering $x_1\rightarrow D$ as a set of data, 
and $x_2\rightarrow \theta$ as the parameter vector of said hypothesis, we can rewrite 
the above equation as
\begin{equation}\label{BayesT}
P(\theta|D, H)=\frac{P(D|\theta,H)P(\theta|H)}{P(D|H)}.
\end{equation}

This last relation is the so-called \textbf{Bayes' theorem} and the most important tool 
in a Bayesian inference procedure. In this result, $P(\theta|D, H)$ is called the
\textbf{{posterior}} probability of the model. $L(D|\theta,H) \equiv P(D|\theta,H) $ 
is called the \textbf{likelihood}, and it will be our main focus in future sections,
$\pi(\theta) \equiv P(\theta|H) $ is called the \textbf{prior} and expresses the knowledge 
about the model before acquiring the data (this prior can be fixed depending on either previous experiment results or the theory behind), $\mathcal{Z} \equiv P(D|H) $ is the evidence of the model, usually referred to as the \textbf{Bayesian Evidence}. 
%MDPI: Please confirm whether the bold are necessary.
% Authors: They're not necessary, but we'd like to stress out these important concepts.

\begin{figure}[t!]
\begin{center}
 \includegraphics[trim = 0mm  0mm 0mm 1mm, clip, width=6.5cm]{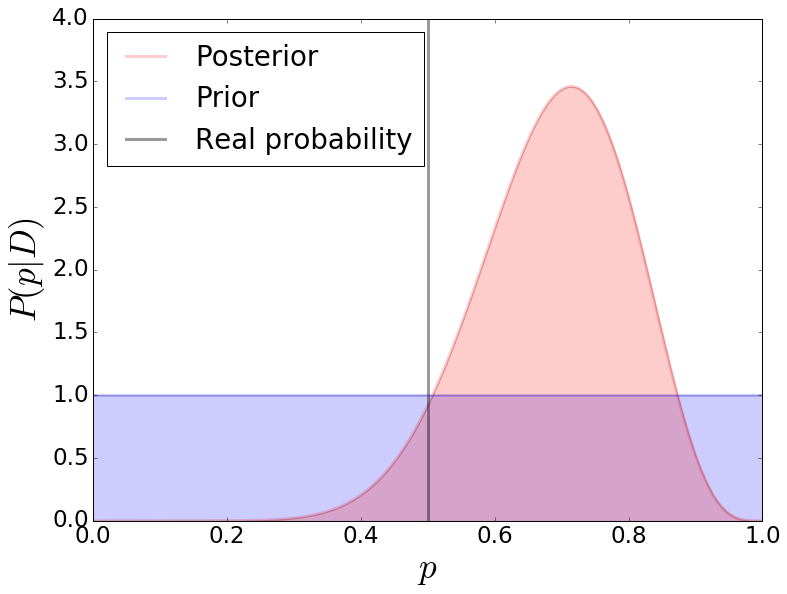} 
\end{center}
\caption{\footnotesize{The coin example: blue figure displays the prior distribution $P(p)$ 
which is updated, when the data is taken into account, to get the posterior distribution $P(p|D)$ (red). 
The vertical black line corresponds to the real value, $p=0.5$.}}
\label{coin0}
\end{figure}

The prior refers to the information one has a priori of the model. It can be defined in various ways; however, a common one is the uniform prior (also referred to as a flat prior):
\begin{equation}\label{flat_prior}
\pi(\theta)\propto c, 
\end{equation}
with $c$ being a constant. This type of prior is telling us that every parameter value is equally probable a priori, as seen in Figure \ref{coin0}. Using this prior also means that the posterior probability will be proportional to the likelihood (since the Bayesian Evidence is a constant). Another convenient prior distribution is the \textit{beta distribution} $B(\theta;a,b)$ since it contains several statistical distributions by varying its parameters $a$ and $b$ (in particular the flat prior is obtained when $a=b=1$). It is defined as
\begin{equation}\label{ex3}
B(\theta;a,b)=\frac{\Gamma(a+b)}{\Gamma(a)\Gamma(b)}\theta^{a-1}(1-\theta)^{b-1},
\end{equation}
where $\Gamma$ is the gamma function. These are just two examples of useful priors, and it is evident that the choice of a prior will influence the posterior distribution, although its effect is reduced as more data are collected, as we shall see later in this section.

Now, regarding the Bayesian Evidence, we notice that it acts as a normalizing factor, and is nothing more than the average of the likelihood over the 
\begin{equation}\label{PD}
P(D|H)=\int d^N\theta P(D|\theta,H)P(\theta|H),
\end{equation}
where $N$ is the dimensionality of the parameter space.
This quantity is usually ignored, for practical reasons, i.e., when testing the parameter 
space of a unique model.
Nevertheless, the Bayesian evidence plays an important role for selecting 
the model that best describes the data, this process being known as \textit{model selection}. 
For convenience, the ratio of two~evidences

\begin{equation}\label{bayesfactor}
K\equiv \frac{P(D\vert H_0)}{P(D\vert H_1)}= 
\frac{\int d^{N_0}\theta_0~P(D\vert\theta_0, H_0) P(\theta_0\vert H_0)}
     {\int d^{N_1}\theta_1~P(D\vert\theta_1, H_1) P(\theta_1\vert H_1)}=
     \frac{\mathcal{Z}_0}{\mathcal{Z}_1},
\end{equation}
or equivalently the difference in $\log$ evidence $\ln \mathcal{Z}_0 - \ln \mathcal{Z}_1$
is often termed as the \textbf{{Bayes factor}} $\mathcal{B}_{0,1}$:

\begin{equation} \label{eq:bayesian}
    \mathcal{B}_{0,1}=\ln  \frac{\mathcal{Z}_0}{\mathcal{Z}_1},
\end{equation}
where $\theta_i$ is a parameter vector (with dimensionality $N_i$) 
for the hypothesis $H_i$ and $i=0,1$. 
In Equation (\ref{eq:bayesian}), the quantity $\mathcal{B}_{0,1}=\ln K$ provides an idea on how 
well model $0$ may fit the data when compared to model $1$. Jeffreys provided a suitable
guideline scale on which we are able to make qualitative conclusions 
(see Table \ref{tab:Jeffrey} \cite{Vazquez:2011xa}).

\begin{table}[t!]
\begin{center}
\begin{tabular}{cccc} 
\cline{1-4}\noalign{\smallskip}
\vspace{0.2cm}
$| \mathcal{B}_{0,1}|$ & Odds & Probability & \,\,\, Strength\\

\hline
\vspace{0.2cm}
$<$ 1.0 &  $< $ 3 : 1  		& $<$ 0.750   & Inconclusive \\
\vspace{0.2cm}
1.0-2.5       &   $\sim$ 12 : 1      		   & 0.923           & Significant \\
\vspace{0.2cm}
2.5-5.0      &    $\sim$ 150 : 1   		    & 0.993           & Strong \\
\vspace{0.2cm}
$>$ 5.0      &       $>$ 150 : 1  	   & $>$ 0.993           & Decisive\\
\hline
\hline
\end{tabular}
\caption{\footnotesize{Jeffreys guideline scale for evaluating the strength of 
evidence when two models are compared.}}
\label{tab:Jeffrey}
\end{center}
\end{table}

We can see that Bayes' theorem has an enormous implication with respect to a statistical
inferential point of view. In a typical scenario, we collect some data and then 
interpret it with a given model; however, we usually do the opposite. That is, first we 
have a set of data, and then we confront a model considering the probability that our 
model fits the data. 
Bayes' theorem provides a tool to relate both scenarios. Then, based on the Bayes' theorem, 
 we are able to select the model that best fits the data.

\textit{Example}.- {We go back to} the example shown in the last section: the coin toss. 
We are interested in the probability of obtaining two heads in a row given the data 
$P(2heads|D)$ ($D$ being the previous 14 coin tosses acting as data). 
First, let us assume that we have a model with a parameter $p$ that
defines the probability of obtaining the two heads, that is $P(2heads|p)$. 
This parameter $p$ will have a probability distribution $P(p|D)$ depending on the data 
in place. Therefore, the probability can be obtained by 
averaging over all the possible parameters with its corresponding density distribution
\begin{equation}\label{ex}
P(2heads|D)=\int^1_0 P(2heads|p)P(p|D)dp.
\end{equation}

For simplicity, we do not update $p$ between the two tosses, but we assume that both of them are 
independent of each other. With this last assumption, we have
\begin{equation}
P(2heads|p)=[P(head|p)]^2 ,
\end{equation}
where $P(head|p)$ is the probability of obtaining a head given our model. 
We assume a simple description of $P(head|p)$ as
\begin{equation}\label{ex1}
P(head|p)=p \quad \Rightarrow \quad P(2heads|p)=p^2.
\end{equation}

On the other hand, notice that we do not know a priori the quantity 
$P(p|D)$ but $P(D|p)$ 
(i.e., we know the probability of obtaining a dataset by considering a model as correct). 
A good choice for experiments that have two possible outcomes is the binomial distribution
\begin{equation}\label{ex2}
    P(x|p,n)=\binom{n}{x}p^{x}(1-p)^{n-x},
\end{equation}
with $n$ the number of trials (in this case, = 14) and $x$ the number of successes (here =10). 
Hence, we have an expression for $P(D|p)$. 
 Using the Bayes' formula, we have
\begin{equation}
P(p|D)=\frac{P(D|p)P(p)}{P(D)}.
\end{equation}

For the prior, we will use the beta distribution \eqref{ex3}, so

\begin{equation}\label{ex4}
P(p)=B(p;a,b).
\end{equation}

In order to get the explicit form of $P(p|D)$, we still need to compute $P(D)$. 
That is, by plugging Equations~\eqref{ex2} and \eqref{ex4} into the integral of Equation~\eqref{PD} yields to
\begin{equation}
P(D)=B(10+a,4+b)\equiv \frac{\Gamma(10+a)\Gamma(4+b)}{\Gamma((10+a)+(4+b))};
\end{equation}
therefore, 
\begin{equation}\label{ex5}
P(p|D)=\frac{p^{10+a-1}(1-p)^{4+b-1}}{B(10+a,4+b)}.
\end{equation}

If we know nothing about $p$, 
then we can assume the prior is a uniform distribution; this means $a=b=1$. 
Notice from Figure \ref{coin0} that our posterior result (Red figure) described by 
Equation \eqref{ex5} does not exactly agree with the real value of $p$ (black dashed vertical line). 
We would expect the posterior distribution to be centered at $p=0.5$ with a very narrow distribution. 
Nevertheless, this value is recovered by increasing the experimental data.

Finally, solving the integral in Equation \eqref{ex} using \eqref{ex1} and \eqref{ex5}, 
we arrive at the result obtained in the previous section
\begin{equation}
P(2heads|D)=\frac{B(13,5)}{B(11,5)}=0.485.
\end{equation}

%\end{Example}
%Authors: the example ends here.
It is important to clarify that the inferred value for the parameter $p$ is merely the probability of said value given our data $D$. This value can change and will generally be a better estimation if more data are collected (as will be seen in detail in the next section).

%
%%%....................................................................................%%%
%%%....................................................................................%%%

%%%------------------------------------------------------------------------%%%
\subsection{Updating the probability distribution}
\label{sub:updating}
%%%------------------------------------------------------------------------%%%

\begin{figure}[t!]
\begin{center}
 \includegraphics[width=8.7cm]{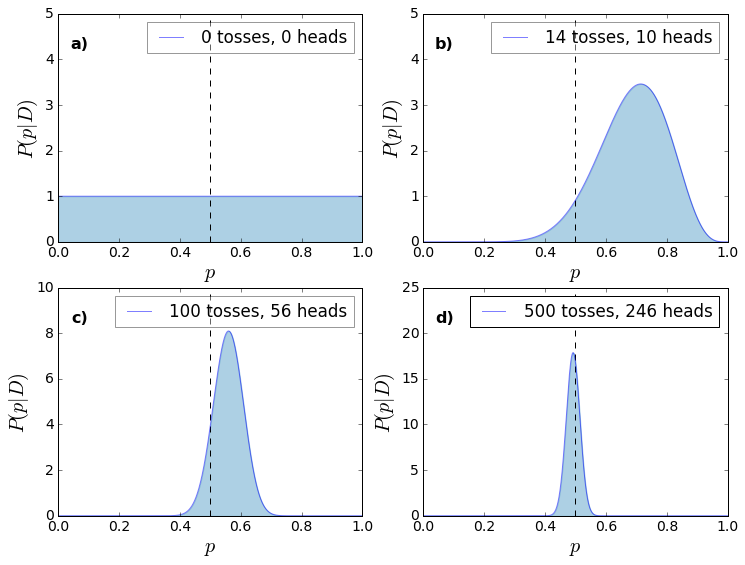}
\end{center}
\caption{\footnotesize{Posterior distributions $P(p|D)$, when the data is increased. 
Notice that while we continue increasing the experimental results, the posterior 
distribution starts to be more localized near by the 
real value $p=0.5$.}}
\label{coin1}
\end{figure}

As seen in the coin example, we were not able to get the real value of $p$ because of the 
lack of enough data. If we want to get closer, we would have to keep 
flipping the coin until the amount of data becomes sufficient. 
Let us continue with the example: 
suppose that after throwing the coin 100 times we obtain, let us say 56 heads, while after 
throwing it 500~times, we obtain 246 heads. Then, we expect to obtain a thinner 
distribution with its center close to $p=0.5$ (see Figure \ref{coin1}). Given this, it is 
clear that in order to confront a parameter and be more accurate about the most 
probable (or ``real'') value, it is necessary to increase the amount 
of data (and the precision) in any experiment. That is, if we take into account the 
500 tosses---with 246 heads---the previous result is updated to $P(2heads|D) = 0.249$,
much closer to the real value.

Then, we have some model parameters that have to be confronted with different sets of data. 
This can be done in two alternative ways: (a) by considering the set of all datasets; 
or (b) by taking each dataset as the new data, but our prior information updated 
by the previous information. The important point in Bayesian statistics is 
that it is indeed equivalent to choose any of these two possibilities. In the coin toss example, 
it means that it is identical to start with the prior given in Figure \ref{coin1}a, and 
then, by considering the 500~data points, we can arrive at the posterior in 
Figure \ref{coin1}d, or similarly 
start with the posterior shown in Figure \ref{coin1}c as our prior and consider only the 
last 400 data points to obtain the same posterior, displayed in Figure \ref{coin1}d. 

%%%....................................................................................%%%
%%%....................................................................................%%%

In fact, if we rewrite Bayes' theorem so that all probabilities are explicitly dependent on 
some prior information $I$ \cite{AlanH}
\begin{equation}\label{BayesTI}
P(\theta|DI,H)=\frac{P(\theta|I,H)P(DI|\theta,H)}{P(D|I,H)},
\end{equation}
and then consider a new set of data $D'$, letting the old data become part of the prior 
information $I'=DI$, we arrive at 
\begin{eqnarray} 
P(\theta|D'I',H) &=&\frac{P(\theta|I,H)P(DD'I|\theta,H)}{P(DD'|I,H)} \nonumber
               \\&=&P(\theta|[DD']I, H),
\end{eqnarray}
where we can explicitly see the equivalence of the two different options. 

%%%------------------------------------------------------------------------%%%
\subsection{About the Likelihood}
\label{sub:likeli}
%%%------------------------------------------------------------------------%%%

We mentioned that the Bayesian evidence is usually set apart when doing any 
inference procedure in the parameter space of a single model. 
Then, without loss of generality, we can fix it to $P(D|H)=1$. 
If we ignore the prior (a motivation for doing this will be expanded upon in the next section), we can identify the posterior with the likelihood 
$P(\theta|D, H)\propto L(D|\theta,H)$; thus, by maximizing it, we can find the 
most probable set of parameters for a model given the data. However, 
having ignored $P(D|H)$ and the prior, 
we are not able to provide an absolute probability 
for a given model, but only relative probabilities. On the other hand, it is 
possible to report results independently of the prior by using the 
\textit{Likelihood ratio}. The likelihood at a particular point in the parameter 
space can be compared with the best-fit value, or the maximum likelihood $L_{\rm max}$. 
Then, we can say that some parameters are acceptable if the likelihood ratio
\begin{equation}
\label{likelihoodratio}
\Lambda=-2\ln\left[\frac{L(D|\theta, H)}{L_{\rm max}}\right],
\end{equation}
is bigger than a given value.

Let us assume we have a single-peaked distribution. 
We consider that $\hat \theta$ is the \textbf{{mean}} of the distribution 
\begin{equation}
\hat \theta =\int d\theta \theta P(\theta|D, H).
\end{equation}

If our model is well-specified and the expectation value of $\hat \theta$ corresponds to the real or most probable value $\theta_0$, we have 
\begin{equation}
\langle\hat \theta\rangle=\theta_0,
\end{equation}
then we say that $\hat \theta$ is \textit{unbiased}. Considering a Taylor expansion of the 
\textit{log likelihood} around its maximum, we have 
{
\begin{eqnarray}
\ln L(D|\theta)&=&\ln L(D|\theta_0) \nonumber  \\ 
&+&\frac{1}{2}(\theta_i-\theta_{0i})\frac{\partial^2\ln L}{\partial\theta_i \partial\theta_j}(\theta_j-\theta_{0j}) \\
&+&..., \nonumber 
\end{eqnarray}
}where $\theta_0$ corresponds to the parameter vector of 
the real model. In this manner, we have that the likelihood can be expressed as a 
multi-variable likelihood given by 
\begin{equation}\label{GLik}
L(D|\theta)=L(D|\theta_0)\exp \small{\left[-\frac{1}{2}(\theta_i-\theta_{0i})H_{ij}(\theta_j-\theta_{0j})\right],}
\end{equation}
where 
\begin{equation}\label{hessianmatrix}
H_{ij}=-\frac{\partial^2\ln L}{\partial\theta_i \partial\theta_j},
\end{equation}
is called the \textbf{{Hessian matrix}}. It controls whether the estimates of 
$\theta_i$ and $\theta_j$ are correlated, and 
if it is diagonal, these estimates are uncorrelated.

The above expression for the likelihood is a good approximation as long as our posterior 
distribution possesses a single-peak. It is worth mentioning that, if the data errors 
are normally distributed, 
then the likelihood for the data will be a Gaussian function as well. In fact, 
this is always true if the model is linearly dependent on the parameters. 
On the other hand, if the data is not normally distributed, we can resort to the 
central limit theorem. In this way, the central limit theorem tells us that 
the resulting distribution will be best approximated by a multi-variate Gaussian 
distribution \cite{LiV}.

%%%------------------------------------------------------------------------%%%
\subsection{Letting aside the priors}
\label{sub:just}
%%%------------------------------------------------------------------------%%%

In this section, we present an argument for letting aside the prior in the parameter estimation. 
For this, we follow the example given in Reference \cite{RobT}. In this example, there are two people, 
A and B, that are interested in the measurement of a given physical quantity $\theta$. 
A and B have different prior beliefs regarding the possible value of $\theta$. 
This discrepancy could be given by the experience, such as the possibility that A and B 
have made a similar measurement at different times. Let us denote their priors by 
$P(\theta|I_i)$, $(i=A,B)$, and assume they are described by two Gaussian distributions 
with mean $\mu_i$ and variance $\Sigma_i^2$. Now, A and B measure 
$\theta$ together using an apparatus subject to a Gaussian noise with known variance $\sigma$. 
They obtain the value $\theta_0=m_1$. Therefore, they can write their likelihoods 
for $\theta$ as %
\begin{equation}\label{LikG}
L(D|\theta, HI)=L_0\exp\left[-\frac{1}{2}\frac{(\theta-m_1)^2}{\sigma^2}\right].
\end{equation}

By using the Bayes formula, the posterior of the model A (and B) becomes
\begin{equation}
P(\theta|m_1)=\frac{L(m_1|\theta I_i)P(\theta|I_i)}{P(m_1|I_i)},
\end{equation}
where we have skipped writing explicitly the hypothesis $H$ and used the notation given in 
Equation~\eqref{BayesTI}. Then, the posterior of A and B are (again) Gaussian with mean
\begin{equation}
\hat \mu_i = \frac{m_1+(\sigma/\Sigma_i)^2\mu_i}{1+(\sigma/\Sigma_i)^2},
\end{equation}
and variance 
\begin{equation}
%\begin{eqnarray}
\tau_i^2=\frac{\sigma^2}{1+(\sigma/\Sigma_i)^2}, \ \ (i=A,B).
%\end{eqnarray}
\end{equation}

\begin{figure}[t!]
\begin{center}
 \includegraphics[trim = 0mm  0mm 0mm 1mm, clip, width=8.cm, height=3.5cm]{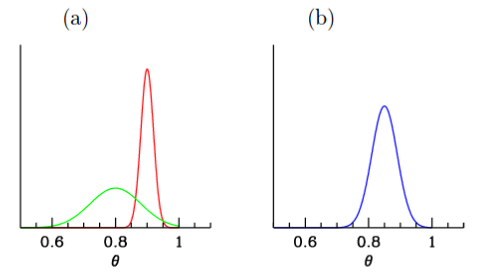}
  \includegraphics[trim = 0mm  0mm 0mm 1mm, clip, width=8.cm, height=3.5cm]{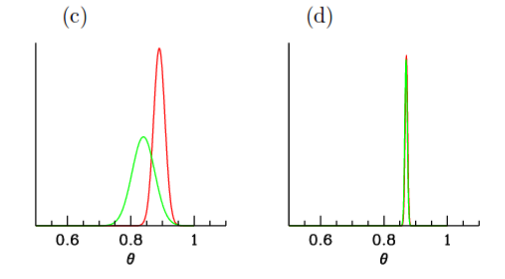}
\end{center}
\caption{\footnotesize{Converging views in Bayesian inference (taken from \cite{RobT}). 
A and B have different priors $P(\theta|I_i)$ for a value $\theta$ (panel (a)). 
Then, they observe one datum with an apparatus subject to a Gaussian noise 
and they obtained a likelihood $L(\theta;HI)$ (panel (b)), after which 
their posteriors $P(\theta|m_1)$ are obtained (panel (c)). After observing 100 data, 
it can be seen that both posteriors are practically indistinguishable (panel (d)).}}
\label{gausian1}
\end{figure}

Thus, if the likelihood is more informative than the prior, i.e., $(\sigma/\Sigma_i)\ll 1$,
the posterior mean of A (and B) will converge towards the measured value $m_1$. As more 
data are obtained, one can simply replace the value of $m_1$ in the above equation by 
the mean $\langle m\rangle$ and $\sigma^2$ by $\sigma^2/N$. 
Then, we can see that the initial prior $\mu_i$ of A and B will progressively be overridden 
by the data. This process is illustrated in Figure \ref{gausian1} where the green (red) 
curve corresponds to the probability distribution of $\theta$ for person A (B) and the 
blue curve corresponds to their~likelihood.

%%%....................................................................................%%%
%%%....................................................................................%%%

%%%------------------------------------------------------------------------%%%
\subsection{Chi-square and goodness of fit}
%%%------------------------------------------------------------------------%%%

We mentioned that the main aim of parameter estimation is to maximize the likelihood 
in order to obtain the most probable set of model parameters given the data. 
If we consider the Gaussian approximation given in Equation \eqref{GLik}, we can see the 
likelihood will be maximum if the quantity
\begin{equation}\label{chi2}
\chi^2\equiv(\theta_i-\theta_{0i})H_{ij}(\theta_j-\theta_{0j}),
\end{equation}
is minimum. The quantity $\chi^2$ is usually called \textbf{{chi-square}} and is related 
to the Gaussian likelihood via $L=L_0e^{-\chi^2/2}$. Then, we can say that 
maximizing the Gaussian likelihood is equivalent to minimizing the chi-square. 
However, as we mentioned before, there are some circumstances where the likelihood 
cannot be described by a Gaussian distribution, in these cases the chi-square 
and the likelihood are no longer equivalent. 

The probability distribution for different values of $\chi^2$ around its minimum is given 
by the $\chi^2$ distribution for $v=n-M$ degrees of freedom, where $n$ is the number of
independent data points and $M$ the number of parameters. Hence, we can calculate the 
probability that an observed $\chi^2$ exceeds, by chance, a value $\hat \chi$ for the 
correct model.
This probability is given by $Q(v,\hat\chi)=1-\Gamma(v/2,\hat\chi/2)$ \cite{NR}, where 
$\Gamma$ is the incomplete Gamma function. Then, the probability that the observed $\chi^2$ 
(even the correct model) is less than a given value $\hat\chi^2$ is $1-Q$. This statement 
is strictly true if the errors are Gaussian and the model is a linear function of the 
likelihood, i.e., for Gaussian likelihoods.

If we evaluate the quantity $Q$ for the best-fit values (minimum chi-square), we can 
have a measure of the goodness of fit. If $Q$ is small (small probability), we can interpret 
it~as:
\begin{itemize}
\item The model is wrong and can be rejected.
\item The errors are underestimated.
\item The error measurements are not normally distributed.
\end{itemize}

On the other hand, if $Q$ is too large there are some reasons to believe that:
\begin{itemize}
\item Errors have been overestimated.
\item Data are correlated or non-independent.
\item The distribution is non-Gaussian.
\end{itemize}

%%%------------------------------------------------------------------------%%%
\subsection{Contour plots and confidence regions}
%%%------------------------------------------------------------------------%%%

Once the best fit parameters are obtained, we would like to know the confidence 
regions where values could be considered good candidates for our model. 
The most logical election is to take values inside a compact region around the best 
fit value. Then, a natural choice are regions with constant $\chi^2$ boundaries. 
When the $\chi^2$ possesses more than one minimum, it is said that we have 
non-connected confidence regions, and for multi-variate Gaussian distributions 
(as the likelihood approximation in Equation \eqref{GLik}) these are ellipsoidal regions. 
In this section, we exemplify how to calculate the confidence regions, 
following Reference \cite{LiV}. 

We consider a small perturbation from the best fit of chi-square 
$\Delta\chi^2=\chi^2-\chi^2_{\rm best}$. Then, we use the properties of $\chi^2$ 
distribution to define confidence regions for variations on $\chi^2$ to its minimum. 
In Table \ref{tableerrors}, we see the typical $68.3 \%$, $95.4\%$, and $99.73\%$ 
confidence levels as a function of number of parameters $M$ for the joint confidence 
level. For Gaussian distributions, these correspond 
to the conventional 1, 2, and 3 $\sigma$ confidence levels. As an example, we plot in {Figure \ref{posteriord}} the corresponding confidence regions associated with the coin~example.
%

%%%....................................................................................%%%
\begin{table}[t!]
\begin{center}
\begin{tabular}{ccccc} 
\cline{1-5}\noalign{\smallskip}
\vspace{0.2cm}
&& & $\Delta\chi^2$ & \\
\quad $\sigma$ \quad & \quad $p$ \quad & \quad  $M=1$ \quad & \quad  $M=2$ \quad & \quad $M=3$ \quad\\

\hline
\vspace{0.1cm}
$1$ & $68.3 \%$ & $1.00$ & $2.30$ & $3.53$\\
\vspace{0.1cm}
$2$ & $95.4 \%$ & $4.00$ & $6.17$ & $8.02$\\
\vspace{0.1cm}
$3$ & $99.73\%$ & $9.00$ & $11.8$ & $14.20$\\
\hline
\end{tabular}
\caption{\footnotesize{$\Delta \chi^2$ for the conventional 
$68.3\%$, $95.4\%$ and $99.73\%$ as a function of the number of parameters ($M$) 
for the joint confidence level.}}\label{tableerrors}
\end{center}
\end{table}
%%%....................................................................................%%%

The general recipe to compute constant $\chi^2$ confidence regions is as follows: 
after finding the best fit by minimizing $\chi^2$ (or maximizing the likelihood) 
and checking that $Q$ is acceptable for the best parameters, then:
\begin{enumerate}
\item Let $M$ be the number of parameters, $n$ the number of data and $p$ the 
confidence limit~desired.
\item Solve the equation:
\begin{equation}
Q(n-M,min(\chi^2)+\Delta\chi^2)=p.
\end{equation}
\item {Find the parameter region where $\chi^2\leq min(\chi^2)+\Delta\chi^2$. This defines the confidence~region.}
\end{enumerate}

%%%------------------------------------------------------------------------%%%
\subsection{Marginalization}
%%%------------------------------------------------------------------------%%%

It is clear that a model may (in general) depend on more than one parameter. 
However, some of these parameters $\theta_i$ may be of less interest. For example, 
they may correspond to nuisance parameters, like calibration factors, or 
it may be the case that we are interested in only one of the parameter constraints 
rather than the joint of two or more of them simultaneously. 
Then, we \textbf{{marginalize}} over the uninteresting parameters by
\begin{equation}\label{marginalization}
P(\theta_1,...,\theta_j,H|D)=\int d\theta_{j+1}...d\theta_{m}P(\theta,H|D),
\end{equation}
where $m$ is the total number of parameters in our model, and $\theta_1$,...,$\theta_j$ 
denote the parameters we are interested in.

%%%------------------------------------------------------------------------%%%
\subsection{Fisher Matrix}
%%%------------------------------------------------------------------------%%%

Once we have a dataset, it is important to know the accuracy for which we can 
estimate parameters. Fisher suggested a way 70 years ago \cite{Fisher}. 
Let us start by considering again a Gaussian likelihood. As we notice, the 
\textbf{{Hessian matrix}} $H_{ij}$ has information on the parameter errors and their 
covariance. More specifically, when all parameters are fixed except one 
(e.g., the $i$-th parameter), its error is $1/\sqrt{H_{ii}}$. 
These errors are called conditional errors, although they are rarely used.

A quantity to forecast the precision of a model, that arises naturally 
with Gaussian likelihoods, is the so-called \textbf{{Fisher information matrix}}
\begin{equation}
F_{ij}=-\left\langle \frac{\partial^2 \mathcal{L}}{\partial \theta_i \partial \theta_j}\right\rangle,
\end{equation}
where 
\begin{equation}
\mathcal{L}=\ln L.
\end{equation}

It is clear that $F=\langle H\rangle$, where the average is made with observational data. 

As we can see from Equation~\eqref{rule3}, for independent datasets, 
the complete likelihood is the product of the likelihoods, and the Fisher 
matrix is the sum of individual Fisher matrices. 
A pedagogical and easy case is having one-parameter $\theta_i$ with a 
Gaussian likelihood. In this~scenario, 

\begin{equation}
\Delta \mathcal{L}=\frac{1}{2}F_{ii}(\theta_i- \theta_{0i})^2,
\end{equation}
when $2\Delta\mathcal{L}=1$, and identifying the $\Delta \chi^2$ corresponding 
to $68\%$ confidence level, we notice that $1/\sqrt{F_{ii}}$ yields the $1-\sigma$ 
displacement for $\theta_i$. In the general case,

\begin{equation}\label{rao}
\sigma_{ij}^2 \geq (F^{-1})_{ij}.
\end{equation}

Thus, when all parameters are estimated simultaneously from the data, the marginalized error is
\begin{equation}
\sigma_{\theta_i}\geq (F^{-1})^{1/2}_{ii}.
\end{equation}

The beauty of the Fisher matrix approach is that there is a simple prescription 
for setting it up by only knowing the model and measurement uncertainties, and under the 
assumption of a Gaussian likelihood the Fisher matrix is the inverse of the 
covariance matrix. So, all we have to do is set up the Fisher matrix and then 
invert it to obtain the covariance matrix (that is, the uncertainties on the model 
parameters). In addition, its fast calculation also enables one to 
explore different experimental setups and optimize the~experiment.

The main point of the Fisher matrix formalism is to predict how well the experiment 
will be able to constrain the parameters, of a given model, before doing the experiment 
and perhaps even without simulating it in any detail. We can then forecast the results of 
different experiments and look at trade-offs, such as precision versus cost. 
In other words, we can engage in experimental design.
The inequality in Equation \eqref{rao} is called the Kramer-Rao inequality. One can see that the 
Fisher information matrix represents a lower bound of the errors. Only when the 
likelihood is normally distributed the inequality is transformed into an equality. 
However, as we saw in Section \ref{sub:likeli}, a Gaussian likelihood is only applicable to some 
circumstances, being generally impossible to be applied, so the key is to have a 
good understanding of our theoretical model in such a way that we can construct 
a Gaussian likelihood.

%%%------------------------------------------------------------------------%%%
\subsubsection{Constructing Fisher Matrices: A simple description}
%%%------------------------------------------------------------------------%%%

Let us construct Fisher matrices in a simple way.
Suppose we have a model that depends on $N$ parameters $\theta_1,\theta_2,...,\theta_N$. 
We consider $M$ observables $f_1,f_2,...,f_M$ each one related to the model parameters 
by some equation $f_i=f_i(\theta_1,\theta_2,...,\theta_N)$. Then, the elements of the 
Fisher matrix can be computed as
\begin{equation}
F_{ij}=\sum_k \frac{1}{\sigma_k^2}\frac{\partial f_k}{\partial\theta_i}\frac{\partial f_k}{\partial\theta_j},
\end{equation}
where $\sigma_k$ are the errors associated with each observable, and we have considered 
they are Gaussianly distributed. 
Here, instead of taking the real data values (which could be unknown), 
it is possible to recreate the data with a fiducial model. 
The errors associated with the mock data can be taken as the expected experimental errors,
and then it is possible to calculate the above expression.

To complement the subject, there is also the \textbf{{Figure of Merit}} 
used by the Dark Energy Task Force (DETF) \cite{detf}, 
which is defined as the reciprocal of the area in the plane enclosing the $95\%$ 
confidence limit of two parameters. The larger the figure of merit the greater accuracy 
one has measuring said parameters. 
%\sout{As an example, let us take a look at \highlighting{Figure 
%\ref{chains_lcdm}} and the lower panel of \highlighting{Figure \ref{LCDM}}, where 
%the area of the error ellipse with only Hubble Data (HD) is clearly bigger 
%than the error ellipse using HD plus several datasets. 
%Then, for this case, the figure of merit would be bigger than with only HD 
%data, since it has a smaller area, making it more accurate for measuring the 
%parameters $\Omega_m$ and $h$.} 
The figure of merit can also be used to see how 
 different experiments break degeneracies, and to predict accuracy in future experiments (experimental design).
%MDPI: The first citation of Figures 15,16,18 are after Figure 4, Figures need to be cited in order, please check the full citation and modify. 
%Authors: Part of the text was removed as it is self-contained in the results, there should be no further problem.

%%%------------------------------------------------------------------------%%%
\subsection{Importance Sampling}
%%%------------------------------------------------------------------------%%%

We call \textbf{{Importance Sampling}} (IS) to different techniques of determining 
properties of a distribution by drawing samples from another one. The main 
idea is that the distribution one samples from should be representative of the distribution 
of interest (for a larger number of samples). In such case, we should infer different 
quantities out of it. In this section, we review the basic concepts necessary to understand 
the IS, following the Reference \cite{importancesampling}.

Suppose we are interested in computing the expectation value $\mu_f=E_p[f(X)]$, 
where $f(X)$ is a probability density of a random variable $X$ and the sub-index $p$ 
means average over the distribution $p$. Then, if we consider a new probability 
density $q(x)$ that satisfies $q(x)>0$ whenever $f(x)p(x)\not = 0$, we can 
rewrite the mean value $\mu_f$ as
\begin{eqnarray}
\mu_f = \int f(x)p(x)dx &=&\int f(x)\frac{p(x)}{q(x)}q(x)dx \nonumber
    \\ &=&E_q[f(X)w(x)],
\end{eqnarray}
where $w(x)=p(x)/q(x)$, and now we have an average over $q$. So, if we have a 
collection of different draws $x^{(1)},...,x^{(m)}$ from $q(x)$, we can 
estimate $\mu_f$ using these draws~as

\begin{equation}
\hat \mu_f = \frac{1}{m}\sum_{j=1}^m w(x^{(j)})f(x^{(j)}).
\end{equation}

If $p(x)$ is known only up to a normalizing constant, the above expression can 
be calculated as a ratio estimate:
\begin{equation}
\hat \mu_f=\frac{\sum_{j=1}^m w(x^{(j)})f(x^{(j)})}{\sum_{j=1}^mw(x^{(j)})}.
\end{equation}

For the strong law of large numbers, in the limit when $m\rightarrow \infty$, 
we will have that $\hat \mu_f\rightarrow \mu_f$.

Another useful quantity to compute in Bayesian analysis is the ratio between 
evidences for two different models
\begin{equation}\label{importanceratio}
\frac{P'(D)}{P(D)}=E\left[\frac{P'(\theta,D)}{P(\theta,D)}\right]_{P(\theta|D)}\simeq \frac{1}{N}\sum_{n=1}^N\frac{P'(D|\theta_n)P'(\theta_n)}{P(D|\theta_n)P(\theta_n)}, 
\end{equation}
where the samples $\lbrace\theta_n\rbrace$ are drawn from $P(\theta|D)$.

An important result for importance sampling is that, if we have a new set of data 
which is broadly consistent with the current data (in the sense that the posterior 
only shrinks), we can make use of importance sampling in order to quickly calculate 
a new posterior including the new data.

%%%------------------------------------------------------------------------%%%
\subsection{Combining datasets: Hyperparameter method}
%%%------------------------------------------------------------------------%%%

Suppose we are dealing with multiple datasets $\lbrace D_1,\ldots,D_N\rbrace$, coming from a 
collection of different surveys $\left\lbrace S_1,\ldots,S_N\right\rbrace$. 
Sometimes it is difficult to know, a priori, if all our data are consistent 
with each other, or whether there could be one or more that are likely to be erroneous. 
If we were sure that all 
datasets are consistent, 
then it should be enough to update the probability, as seen in Section \ref{sub:updating}, in order to 
calculate the new posterior distribution for the parameters we are interested in. However, 
since there is usually an uncertainty about this, a way to know how useful 
the data may be is by introducing the \textbf{{hyperparameter method}}. This method was 
initially introduced by Reference \cite{hiperp, hiperp1} in order to perform a joint 
estimation of cosmological parameters from combined datasets. This method may be used as 
long as every survey is independent of each other.
In this section, we review the main steps necessary to understand the hyperparameter 
method.

The main feature of this process is the introduction of a new set of hyperparameters 
$\alpha$ in the Bayesian procedure to allow extra freedom 
in the parameter estimation. These hyperparameters are equivalent to nuisance parameters in 
the sense that we need to marginalize over them in order to 
recover the posterior distribution, i.e.,
\begin{equation}
P(\theta|D,H)=\frac{1}{P(D|H)}\int P(\theta|\alpha,H)P(\alpha|D, H)d\alpha,
\end{equation}
where we have used the Bayes' theorem. Now, for the method, it is necessary to assume that the 
hyperparameters $\alpha$ and the parameters of interest $\theta$ are independent, i.e.,
$P(\theta,\alpha,H)=P(\alpha)P(\theta,H)$, and it is also necessary to assume that each 
hyperparameter $\alpha_k$ is independent of each other, i.e., 
$P(\alpha)=P(\alpha_1)P(\alpha_2)\ldots P(\alpha_N)$. 
In this way, we can rewrite the above expression as
\begin{equation*}
P(\theta|D,H)=\frac{P(\theta,H)}{P(D|H)}\left[\prod_{k=1}^N\int P(D_k|\theta,\alpha_k,H)P(\alpha_k)d\alpha_k\right].
\end{equation*}

Here, the quantity inside the square brackets is the marginalized likelihood over the
hyperparameters. We can identify the quantity inside 
the integration as the individual likelihood $L(D_k|\theta,\alpha_k,H)$, for every 
$\alpha_k$ and the dataset $D_k$; $P(D|H)$ is the evidence and, similarly to a parameter 
inference procedure, it works as a normalizing function, i.e., 
$P(D|H)=\int d \theta P(\theta,H)L(D|\theta,H)$. 
Notice that, by considering $P(\alpha_k)=\delta(\alpha_k-1)$, we rely on the standard 
approach, where no hyperparameters are used. 

We add these $\alpha_k$ in order to weight every dataset and take away the 
data that does not seem to be consistent with other ones. Then, we would like to 
know whether the data supports the introduction of hyperparameters or not. 
A way to address this point is given by the Bayesian evidence $K$ defined in 
Equation~\eqref{bayesfactor}.
If we consider a Gaussian likelihood with maximum entropy prior, and assuming that 
on average the hyperparameters' weights are unity, we can rewrite the marginalized 
likelihood function $L(D|\theta,H_1)$ for model $H_1$ as
\begin{equation}\label{hyperlik}
P(D|\theta,H_1)=\prod_{k=1}^N\frac{2\Gamma(\frac{n_k}{2}+1)}{\pi^{n_k/2}|V_k|^{1/2}}(\chi_k^2+2)^{-\left(\frac{n_k}{2}+1\right)},
\end{equation}
obtaining an explicit functional form for $K$, given by 
\begin{equation}\label{bayesfactork}
K=\prod_{k=1}^N\frac{2^{n_k/2+1}\Gamma(n_k/2+1)}{\chi^2_k+2}e^{-\chi_k^2/2}.
\end{equation}

Here, $\chi_k^2$ is given by (\ref{chi2}) for every dataset, and $n_k$ is the number of 
points contained in $D_k$. In Equation \eqref{hyperlik}, $V_k$ is the covariance matrix for the $k$-data.
Suppose we have two models, one with hyperparameters, called $H_1$, 
and a second one without them, called $H_0$. The Bayesian evidence $P(D|H_i)$ is the
key quantity for making a comparison between two different 
models. In fact, by using the Bayes factor $K$ from Equation~(\ref{bayesfactork}), we can estimate the 
necessity to introduce the hyperparameters to our model using the criteria given in 
Table \ref{tab:Jeffrey}.
Notice that, if we have a set of independent samples for 
$H_0$, we can compute an estimate for $K$ with the help of Equation \eqref{importanceratio}.

%%%================================================================%%%
\section{Numerical tools}
\label{sec:tools}
%%%================================================================%%%

In typical scenarios, it is very difficult to compute %result of 
%Please ensure the meaning has been retained.
the posterior distribution analytically. 
For these cases, the numerical tools play an important role during 
the parameter estimation task. There exist several options to carry out this work; 
nevertheless, in this section, we focus on the Markov Chain Monte Carlo (\textbf{{MCMC}}) 
with the Metropolis Hastings algorithm (MHA). 
Additionally, we present some useful details we 
take into account to make our computation more efficient . 

%%%------------------------------------------------------------------------%%%
\subsection{MCMC techniques for parameter inference}
%%%------------------------------------------------------------------------%%%

The purpose of the MCMC algorithm is to build up a sequence of points 
(called \textbf{{chain}}) in 
a parameter space in order to evaluate the posterior of Equation \eqref{BayesT}. In this 
section, we review the basic steps for this procedure in a simplistic way; however, 
it is recommendable to check \cite{medel, mcmc1, mcmc2, mcmc3, mcmc4} for 
a more formal version of the {MCMC} theory.
%MDPI: The italics or normal are not consistent, please check the full text and revise
%Authors: We changed the order of the bold text.

A \textbf{{Monte Carlo}} simulation is assigned to algorithms that use random number 
generators to approximate a specific quantity. On the other hand, 
a sequence $X_1,X_2,\ldots$ of elements of some set is a \textbf{Markov Chain} 
if the conditional distribution of $X_{n+1}$ given $X_1,\ldots,X_n$ depends only on $X_n$. 
In other words, a Markov Chain is a process where we can compute subsequent steps 
based only in the information given at the present. An important property of a Markov 
Chain is that it converges to a stationary state where successive elements of the chain 
are samples from the target distribution; in our case, it converges to the posterior 
$P(\theta|D,H)$. Hence, we can estimate all the usual quantities of interest out of the 
posterior (mean, variance, etc.). 

The combination of both procedures is called an MCMC. %\textbf{\hl{MCMC}}. 
The number of points required to get good estimates 
in MCMCs is said to scale linearly with the number of parameters, so this method 
becomes much faster than grids as the dimensionality increases.

The target density is approximated by a set of delta functions:
\begin{equation}
p(\theta|D,H)\simeq \frac{1}{N}\sum_{i=1}^N \delta(\theta-\theta_i),
\end{equation}
with $N$ being the number of points in the chain. 
Then, the posterior mean is computed as
\begin{equation}
\langle\theta\rangle=\int d\theta \theta P(\theta,H|D)\simeq \frac{1}{N}\sum_{i=1}^N\theta_i ,
\end{equation}
where $\simeq$ follows because the samples $\theta_i$ are generated out of 
the posterior by construction. Then, we can estimate any integrals (such as the mean, variance, etc.) as
\begin{equation}
\langle f(\theta)\rangle \simeq\frac{1}{N}\sum_{i=1}^N f(\theta_i).
\end{equation}

As mentioned before, in a Markov Chain, it is necessary to generate a new point $\theta_{i+1}$ 
 from the present point $\theta_i$. However, as it is expected, we need a criterion 
for accepting (or rejecting) this new point depending on whether it turns out to be better for 
our model or not. If the new step is worse than the previous one,
we may still accept it since, if we only accept steps with better 
probability, we could be converging into a local maximum in our parameter space and, 
therefore, just exploring a small region of the entire space. The simplest algorithm that contains all this 
information in its methodology is known as the Metropolis-Hastings algorithm.

%%%------------------------------------------------------------------------%%%
\subsubsection{Metropolis-Hastings algorithm}
%%%------------------------------------------------------------------------%%%

In the \textbf{{Metropolis-Hastings algorithm}} \cite{MHas,Mhas2}, it is necessary to 
start from a random initial point $\theta_i$, with an associated posterior 
probability $p_i=p(\theta_i|D,H)$. We need to propose a candidate $\theta_c$ 
by drawing from a \textbf{proposal distribution} $q(\theta_i,\theta_c)$ 
 used as a generator of new random steps. Then, the probability of 
acceptance of the new point is given by%
\begin{equation}
p(acceptance)=min\left[1,\frac{p_cq(\theta_c,\theta_i)}{p_iq(\theta_i,\theta_c)}\right].
\end{equation}

If the proposal distribution is symmetric, the algorithm is reduced to the 
\textit{Metropolis algorithm}
  \begin{equation}
  p(acceptance)=min\left[1,\frac{p_c}{p_i}\right].
  \end{equation}

In this way, the complete algorithm can be expressed by the following steps:
\begin{enumerate}
\item Choose a random initial condition $\theta_i$ in the parameter space and compute 
the posterior~distribution.
\item Generate a new candidate from a proposal distribution in the parameter space and compute 
the corresponding posterior distribution.
\item Accept (or reject) the new point with the help of the Metropolis-Hastings algorithm.
\item If the point is rejected, repeat the previous point in the chain.
\item Repeat steps 2--4 until you have a large enough chain.
\end{enumerate}

%%%------------------------------------------------------------------------%%%
\subsubsection{A first example of parameter inference}
%%%------------------------------------------------------------------------%%%

\begin{figure}[t!]
\begin{center}
 \includegraphics[trim = 0mm  0mm 0mm 0mm, clip, width=6.5cm]{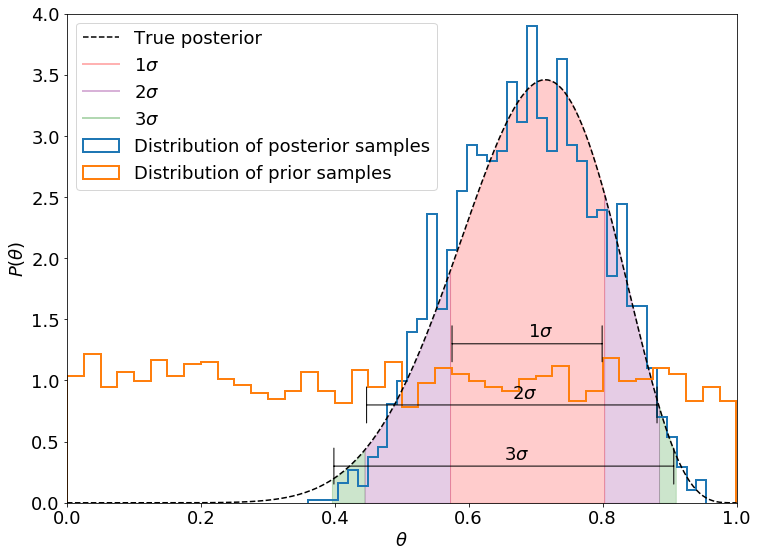}
\end{center}
\caption{\footnotesize{ 1D posterior distribution for our example. 
We plot the prior distribution (red), true posterior (dashed-black) and 
the posterior calculated by the MHA (blue). We plot 1,2 and 3$\sigma$ confidence regions 
for the estimation of $p$.}}
\label{posteriord}
\end{figure}

In order to put into practice the numerical tools, we take again 
the coin toss example seen from Section \ref{sub:BTPP}. 
Let us try to estimate the value of $p$ (or region of values for $p$) 
that best matches our data (we assume only the 14 times that the coin was thrown). 
To calculate the posterior distribution \eqref{ex4}, we use the MHA. 

As mentioned before, we consider a likelihood given by the binomial distribution 
\eqref{ex2} and a normal distributed prior \eqref{ex3} ($a=b=1$). 
As our first guess for $p$, we consider $p_i=0.1$. We generate a new 
candidate $p_c$ as $p_c=p_{cu}+G(p_{cu},\hat\sigma)$, where 
$G(p_{cu},\hat\sigma)$ is our proposed Gaussian distribution centered 
at $p_{cu}$ with variance $\hat\sigma=0.1$; $p_{cu}$ is the current 
value of $p$, for our first step is $p_{cu}=p_i$. Then, we introduce the 
MHA in a Python code. 
Our final result (shown in Figure \ref{posteriord}) is a posterior distribution 
that matches very well with the results calculated analytically (shown in Figure \ref{coin0}). 
Numerically, we obtained $p= 0.695^{+0.123}_{-0.107}$, where the upper and lower 
values for $p$ correspond to the $1\sigma$ standard deviation.
Notice that we have plotted the width of our $1\sigma$, $2\sigma$ 
and $3\sigma$ confidence regions in the same figure. 
To complement the example (and Figure \ref{posteriord}), we also show in Figure \ref{chain2} 
the Markov Chains generated by our code. 
It is easy to see that the chains oscillate with small amplitude 
around the mean value.

\begin{figure}[h!]
\includegraphics[width=1.1\textwidth]{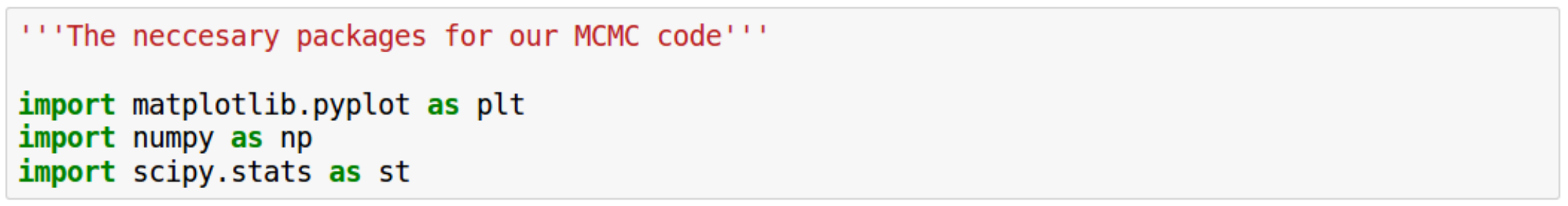}
\includegraphics[width=1.1\textwidth]{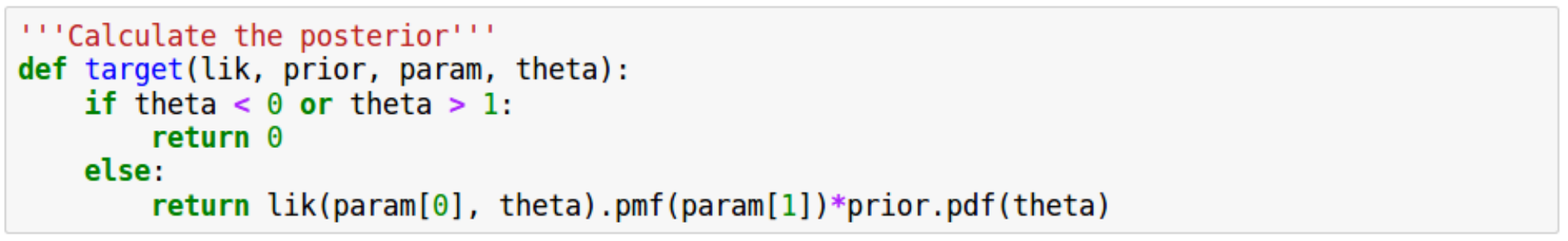}
\includegraphics[width=1.1\textwidth]{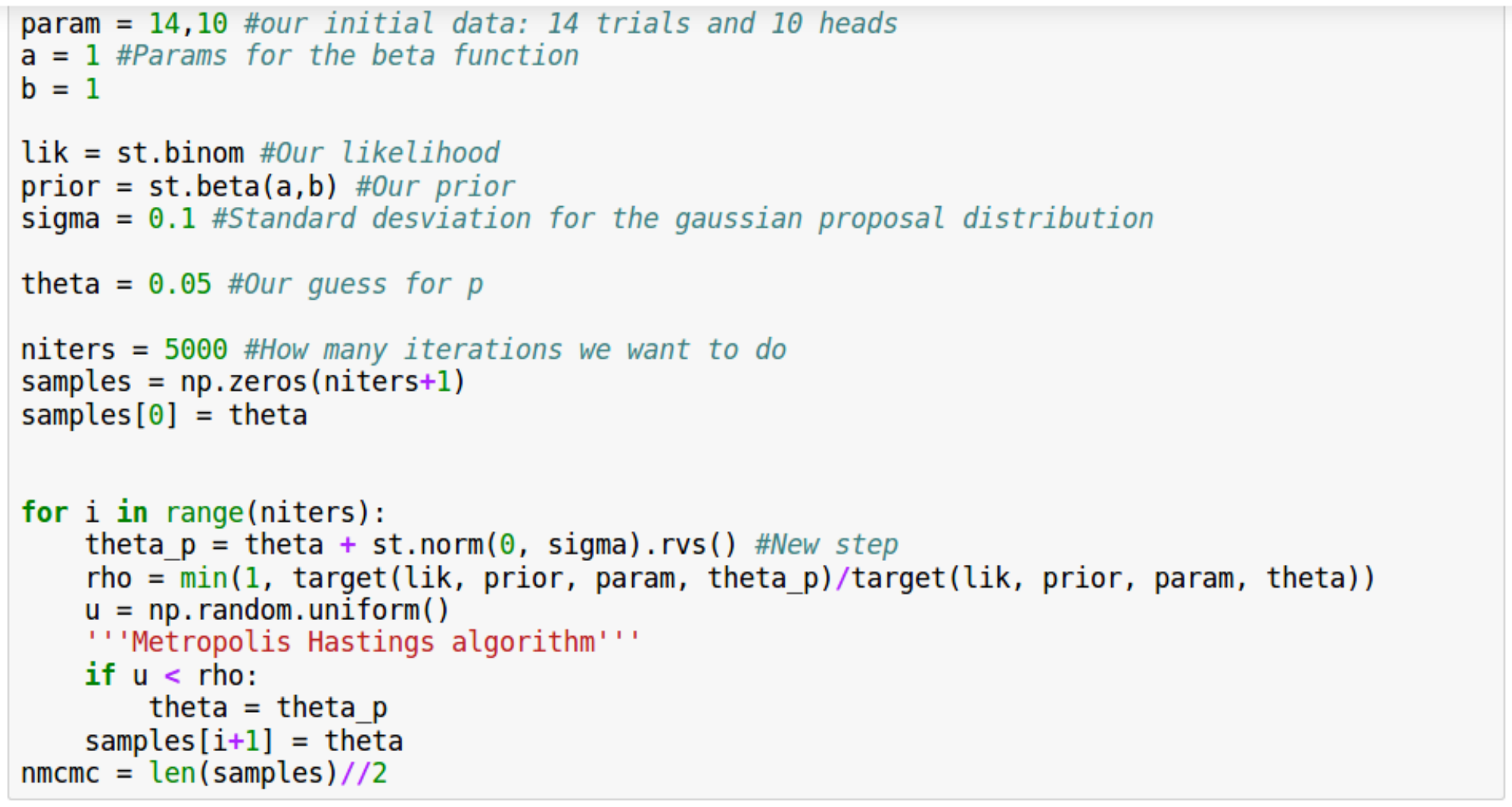}
\includegraphics[width=1.1\textwidth]{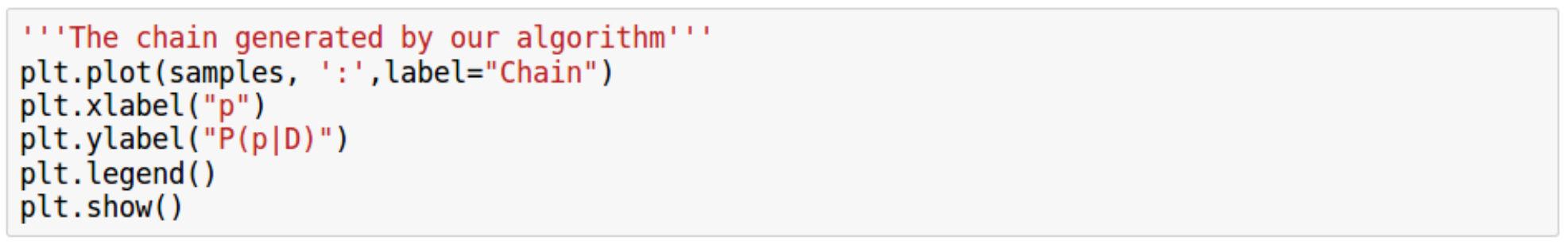}
\caption{\footnotesize{Here we show our MCMC code, written in Python, from scratch.}}
\label{fig:MCMCcode}
\end{figure}

Remark: In Figure \ref{fig:MCMCcode}, we include the MCMC algorithm using an explicit 
code for the MCMC process. However, in Python there are some modules that 
can simplify this task. For example, PyMC3 \cite{pymc3} is a 
Python module that implements statistical models and fitting algorithms, 
including the MCMC algorithm. We use this module at the end of this 
section.

%%%------------------------------------------------------------------------%%%
\subsubsection{Convergence test} 
%%%------------------------------------------------------------------------%%%

It is clear that we need a test to know when our chains have converged. 
We need to verify that the points in the chain are not converging to a 
false convergent point or to a local maximum. In this sense, we need 
that our algorithm takes into account this possible difficulty. The simplest way 
(the informal way) to know whether our chain is converging to a global maximum 
is by running several chains starting with different initial proposals for 
the parameters. Then, if we see by the naked 
eye that all chains seem to converge into a single region of the 
possible value for our parameter, we may say that our chains are 
converging to that region. 

\begin{figure}[t!]
\begin{center}
 \includegraphics[width=7cm]{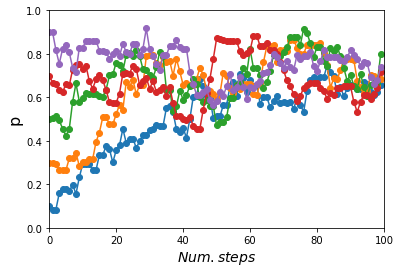}
\end{center}
\caption{\footnotesize{The five Markov 
Chains used to estimate the posterior distribution.  We use $p=0.1,0.3,0.5,0.7,0.9$ for 
starting points.}}
\label{chain2}
\end{figure}

Taking yet again the example of the coins, we can run several chains for 
the above example and try to estimate whether the value (region) of $p$ 
that we found is a stationary value. In Figure \ref{chain2}, 
we plot 5 different Markov chains with initial conditions 
$p=0.1,0.3,0.5,0.7,0.9$. As we expected from the analytical result, 
after several steps all the chains seem to concentrate nearby the same value.

The convergence method used above is very informal, and we would like to have 
a better way to ensure that our result is correct. The usual test 
is the \textit{Gelman-Rubin} convergence criterion 
\cite{Gelm,Gelm2}. 
That is, by starting with $M$ 
chains with very different initial points and $N$ points per chain, 
if $\theta_i^j$ is a point in the parameter space of position $i$ 
and belonging to the chain $j$, we need to compute the mean of each chain:
\begin{equation}
\langle\theta^j\rangle =\frac{1}{N}\sum_{i=1}^N \theta_i^j,
\end{equation}
and the mean of all the chains
\begin{equation}
\langle\theta\rangle =\frac{1}{NM}\sum_{i=1}^N\sum_{j=1}^M\theta_i^j.
\end{equation}

Then, the chain-to-chain variance $B$ is
\begin{equation}
B=\frac{1}{M-1}\sum_{j=1}^M(\langle\theta^j\rangle-\langle\theta\rangle)^2 ,
\end{equation}
and the average variance of each chain is
\begin{equation}
W=\frac{1}{M(N-1)}\sum_{i=1}^N\sum_{j=1}^M(\theta_i^j-\langle\theta^j\rangle)^2 .
\end{equation}

If our chains converge, $W$ and $B/N$ must agree. In fact, we say that the 
chains converge when the quantity
\begin{equation}
\hat R=\frac{\frac{N-1}{N}W+B(1+\frac{1}{M})}{W},
\end{equation}
which is the ratio of the two estimates, approaches unity. A typical convergence 
criterion is when $0.97<\hat R<1.03$. 

%%%------------------------------------------------------------------------%%%
\subsubsection{Some useful details}
%%%------------------------------------------------------------------------%%%

\begin{figure}[t!]
\begin{center}
 \includegraphics[trim = 0mm  0mm 0mm 1mm, clip, width=7.5cm, height=4.5cm]{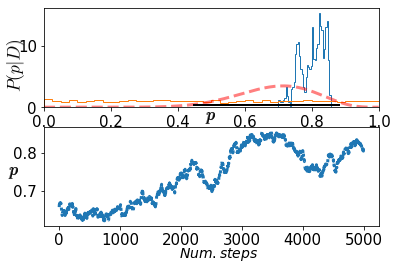}
\includegraphics[trim = 0mm  0mm 0mm 1mm, clip, width=7.5cm, height=4.5cm]{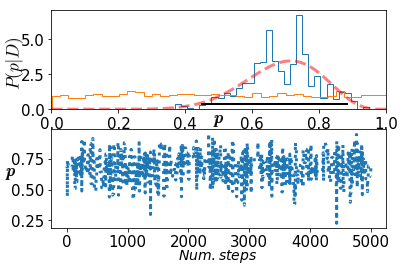}
\end{center}
\caption{\footnotesize{Two Markov Chains considering 
different variance for our Gaussian proposal distribution. 
Upper panel corresponds to $\hat \sigma=0.002$, while lower panel 
corresponds to $\hat \sigma = 0.8$.}}
\label{chainprop}
\end{figure}

%
\textbf{{The proposal distribution:}} The choice of a proposal distribution $q$ 
is crucial for the efficient exploration of the posterior. In our example, 
we used a Gaussian-like distribution with a variance (step) $\hat\sigma=0.1$. 
This value was taken because we initially explored different values for 
$\hat\sigma$, and we select the quickest that
approaches the analytic posterior distribution of $p$. However, if the scale of 
$q$ is too small compared to the scale of the target (in the sense that 
the typical jump is small), then the chain may take very long to 
explore the target distribution, which implies that the algorithm will be 
very inefficient. As we can see in Figure \ref{chainprop} (top panel), 
considering an initial step $p_i=0.6$ and a variance for the proposal 
distribution $\hat\sigma = 0.002$, the number of points 
are not enough for the system to move to its ``real'' posterior 
distribution. On the other hand, if the scale of $q$ is too large, 
the chain gets stuck, and it does not jump very frequently (bottom panel 
of the figure corresponding to $\hat\sigma = 0.8$), so we will have different 
maxima in our posterior. 

In order to fix this issue in a more efficient way, it is recommendable 
to run an exploratory MCMC, compute the covariance matrix from the samples, 
and then re-run with this covariance matrix as the covariance of a 
multivariate Gaussian proposal distribution. This process 
can be computed a couple of times before running the final MCMC.

\textbf{{The burn-in:}} It is important to notice that, at the beginning of the chain, we
have a set of points far outside the stationary region. 
This early stage of the chain (called burn-in) must be ignored; this means 
that the dependence on the starting point must be lost. Thus, it 
is important to have a reliable convergence test.

 \textbf{{Thinning:}} There are several Bayesian statisticians 
 that usually thin their MCMC, which means that they do not prefer to 
 save every step given by the MCMC; instead, they prefer to save a new 
 step each time $n$ steps have taken place. An obvious consequence 
of thinning the chains is that the amount of autocorrelation is 
 reduced. However, as long as the chains are thinned, the precision for the 
 estimated parameters is reduced \cite{thin}. Thinning the chains can be 
 useful in other kind of circumstances, for example, if we have 
 limitations in memory. Notice that thinning a chain does not yield 
 incorrect results; it yields correct results, but less efficient ones, than 
 using the full chains.

\textbf{{Autocorrelation probes:}} A complementary way to look for 
convergence in an MCMC estimation is by looking for the autocorrelation 
between the samples. The autocorrelation $lag\ k$ is defined as the 
correlation between every sample and the sample $k$ steps before. 
It can be quantified as \cite{autocor,autocor2}
\begin{eqnarray}
\rho_k &=&\frac{Cov(X_t,X_{t+k})}{\sqrt{Var(X_t)Var(X_{t+k})}} \nonumber \\ &=&
\frac{E[(X_t-X)(X_{t+k}-X)]}{\sqrt{E[(X_t-X)^2]E[(X_{t+k}-X)^2]}},
\end{eqnarray}
{where $X_{t}$ is the $t$-th sample, and $X$ is the mean of the samples. 
This autocorrelation should become smaller as long as $k$ increases 
(this means that samples start to become~independent).}

\subsubsection*{More Samplers}

\textbf{{Gibbs sampling:}} The basic idea of the Gibbs sampling algorithm
\cite{gibbs} is to split the multidimensional $\theta$ into 
blocks and sample each block separately, conditional on the most recent 
values of the other blocks. It basically breaks a high-dimensional problem 
into low-dimensional problems.

The algorithm reads as follows:
\begin{enumerate}
\item $\theta$ consists of $k$ blocks $\theta_1, \ldots, \theta_k$. Then, at step $i$
\item Draw $\theta^{i+1}_{1}$ from $p(\theta_1|\theta^{i}_{2}, \ldots,\theta^{i}_{k})$
\item Draw $\theta^{i+1}_{2}$ from $p(\theta_2|\theta^{i+1}_{1}, \theta^{i}_{3}, \ldots,\theta^{i}_{k})$
\item \ldots
\item Draw $\theta^{i+1}_{k}$ from $p(\theta_k|\theta^{i+1}_{1}, \theta^{i+1}_{2}, \ldots,\theta^{i+1}_{k-1})$
\item Repeat the above steps for the wished iterations with $i \rightarrow i+1$.
\end{enumerate}

The distribution $p(\theta_1|\theta_{2}, \ldots,\theta_{k})=\frac{p(\theta_1, \ldots, \theta_k)}{p(\theta_2, \ldots, \theta_k)}$ is known as the 
{full conditional distribution of $\theta_1$}. This algorithm 
is a special case of MHA where the proposal is always accepted.
%MDPI; Please confirm whether the italics are necessary. 
%Authors: For 'full conditional', they are not necessary.
%however, we used italics to stress out definitions. 

\textbf{{Metropolis} Coupled Markov Chain Monte Carlo ($MC^3$):} It is easy to see that it could be a little problematic if our likelihood has local maxima. The $MC^3$ is a modification of the standard MCMC algorithm that consists of running several Markov Chains in parallel to explore the target distribution for different temperatures. The temperature controls the height of the peaks in the likelihood; this simplifies 
the way we sample our parameter space and helps us to avoid local maxima. If a distribution has a temperature $T<1$, it is said to be tempered. 

We consider a tempering version of the posterior distribution $P(\theta,T|D,H)$:
\begin{equation}
P(\theta,T|D,H) \propto L(\theta,D)^{1/T}P(\theta,H),
\end{equation}
where $L$ is the likelihood, and $P(\theta,H)$ the prior. Notice that, 
for higher $T$, individual peaks of $L$ become flatter, making the distribution 
easier to sample with an MCMC algorithm. Now, we have to run N chains with 
different temperatures assigned in a ladder $T_1<T_2<\ldots<T_N$, usually 
taken with a geometrically distributed division, with $T_1=1$. The coldest 
chain $T_1$ samples the posterior distribution more accurately 
and behaves as a typical MCMC. Then, we define this chain as the main chain. 
The rest of the chains are running such that they can cross local 
maximum likelihoods easier and transport this information to our main chain. 

The chains explore independently the landscape for a certain number of 
generations. Then, in a pre-determined interval, the chains are allowed 
to swap its actual position with~probability
\begin{equation}
A_{i,j}=min\left\lbrace\left(\frac{L(\theta_i)}{L(\theta_j)}\right)^{1/T_j-1/T_i},1\right\rbrace .
\end{equation}

In this way, if a swap is accepted, chains $i$ and $j$ must exchange 
their current position in the parameter space, and then chain 
$i$ has to be in position $\theta_j$ and chain $j$ has to move to 
position $\theta_i$. 

We can see that, since the hottest chain $T_{\rm max}$ can access easier to all the 
modes of $P(\theta,H,T_{\rm max}|D)$, then it can propagate its position 
to colder chains, to be precise, it can propagate its position to the 
coldest chain $T=1$. At the same time, the position of colder chains 
can be propagated to hotter chains, allowing them to explore the entire prior volume. 
For an extensive explanation, or modification to make the temperature of the chains dynamical, 
see Reference \cite{mcmcmc, PopulationMC, kilbinger2012cosmopmc}.

\textbf{{Affine Invariant} MCMC Ensemble Sampler:} 
{The main property of this algorithm relies on its invariance under affine transformations. 
Let us consider a highly anisotropic~density: }
\begin{equation}
    p(x_1, x_2) \propto \exp{\bigg( \frac{-(x_1-x_2)^2}{2\epsilon}-\frac{(x_1+x_2)^2}{2}\bigg)},
\end{equation}
which is difficult to calculate for small $\epsilon$. But, by making the affine transformation
\begin{equation}
    y_1=\frac{x_1-x_2}{\sqrt{\epsilon}}, \quad y_2=x_1+x_2,
\end{equation}
we can rewrite the anisotropic density into the easier problem 
\begin{equation}
    p(y_1, y_2) \propto \exp{\bigg( \frac{-(y_1^2+y_2^2)}{2}\bigg)}.
\end{equation}

An MCMC sampler has the form $X(t+1)=R(X(t),\psi (t),p)$, where $X(t)$ is the 
sample after $t$ iterations, $R$ is the sampler algorithm, $\psi$ is the sequence of independent identically 
distributed random variables, and $p$ is the density. 
A sampler is said to be affine 
invariant if, for any affine transformation $Ax+b$,
\begin{equation}
    R(AX(t)+b,\psi (t),p_{A,b})=AR(X(t),\psi (t),p)+b.
\end{equation}
There are already several algorithms that are affine invariant, one of the easiest 
is known as the \textit{stretch move} \cite{stretch}. 
An algorithm fully implemented in Python under the name 
 \textbf{{EMCEE}}~\cite{emcee} is also affine invariant, and some others that can be found in Reference \cite{affine}.

\textbf{{Even more samplers:}} The generation of the elements in a Markov chain 
is probabilistic by construction, and it depends on the algorithm we are 
working with. The MHA is the easiest algorithm used in Bayesian inference. 
However, there are several algorithms that can help to accomplish 
our mission, for instance, some of the most popular and effective ones, 
are the Hamiltoninan Monte Carlo (see, e.g.,
Reference \cite{hamiltonian1,Hamiltonian2}) or the Adaptative Metropolis-Hastings 
(AMH) (see, e.g., Reference \cite{importancesampling}).

 %%%------------------------------------------------------------------------%%%
 \section{Fitting a straight-line} \label{sec:line}
%%%------------------------------------------------------------------------%%%

\begin{figure}[t!]
\begin{center}
\includegraphics[trim = 0mm  0mm 0mm 1mm, clip, width=7.5cm, height=4.3cm]{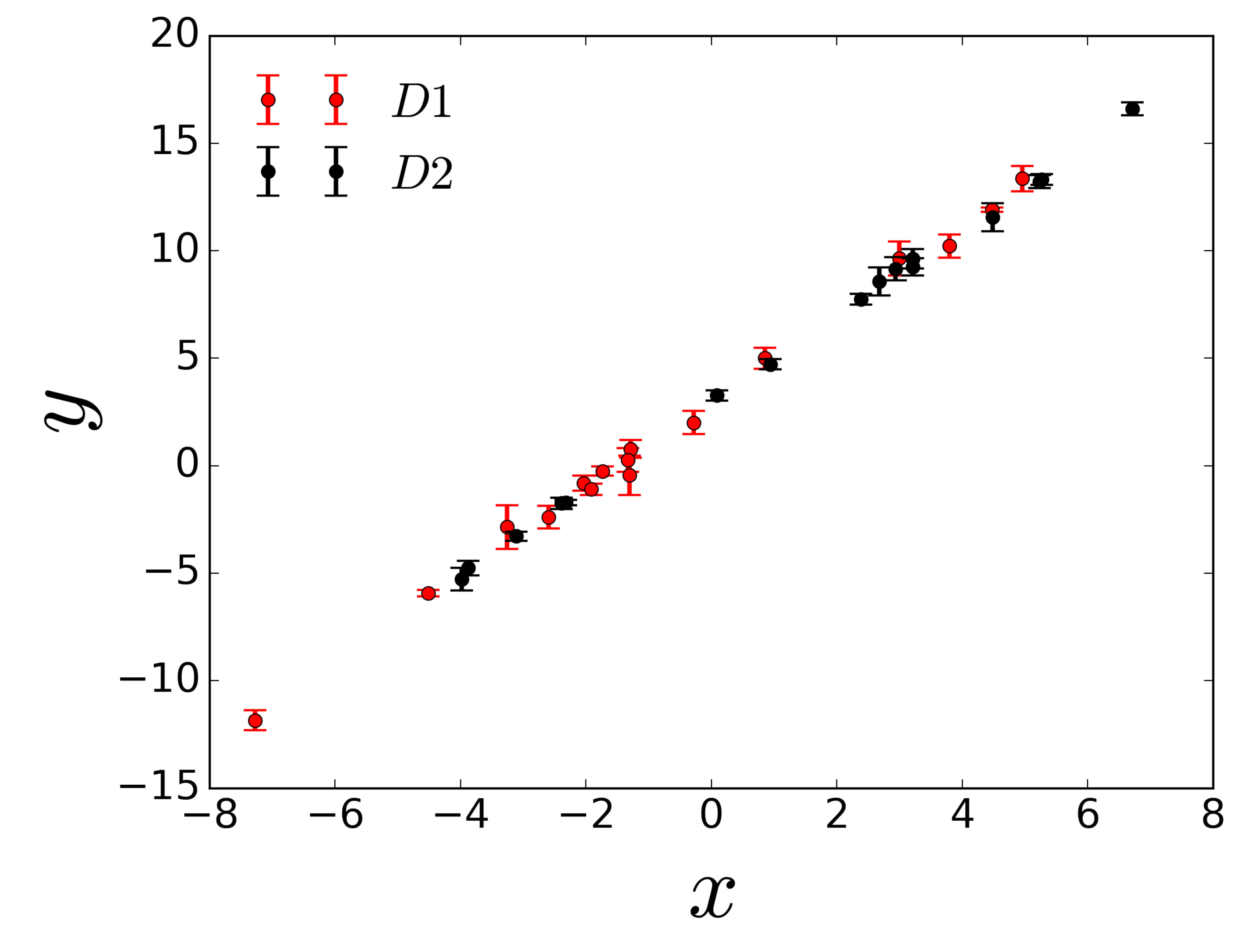} \qquad
\includegraphics[trim = 0mm  0mm 0mm 1mm, clip, width=7.5cm, height=4.3cm]{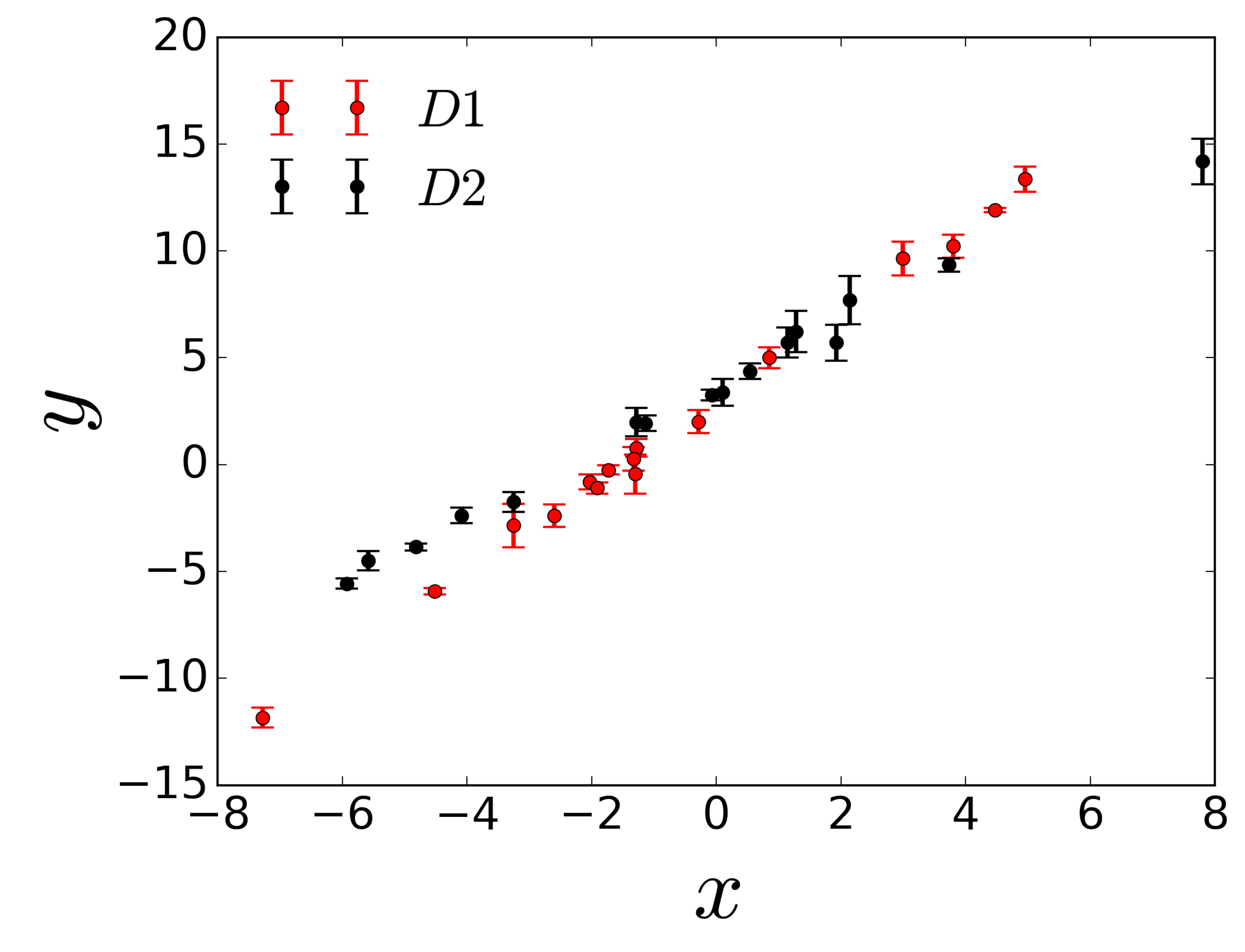}
\end{center}
\caption{\footnotesize{Datasets $D_1$ and $D_2$ measured by the straight-line theory.
Top: case 1 and Bottom: case 2.}}
\label{data_1}
\end{figure}

{In this section, we carry out an standard }
example: fitting a straight-line. That is, we assume that we have a certain 
theory where our measurements should follow a straight line. Then, 
we simulate several datasets and focus on two different~cases (Figure \ref{data_1}):
%MDPI: We have changed it, please confirm,
%Authors: we confirm, however we replaced 'the simplest' by 'an standard'
%
%
\begin{enumerate}
    \item Consider two datasets coming from the same 
straight-line but having different errors.
    \item Consider two datasets coming 
from different straight-lines and also having different~errors.
\end{enumerate}

In our analysis, we used the PyMC3 module, and the complete code can be 
downloaded from the Git repository \cite{git}.

%%%------------------------------------------------------------------------%%%
\subsection{Case 1}
%%%------------------------------------------------------------------------%%%

In this example, we start by considering that our measurements for a given 
theory (a straight-line $y=a+bx$) are 
given by the data shown in the upper panel of Figure \ref{data_1}. 
The two datasets, D1 and D2, were generated from 
the same line $y=3+2x$, adding a gaussian error to each point. For D1, we 
added an error with a standard deviation $\sigma_1 = 0.3$, while, for D2 
we use $\sigma_2 = 0.2$. Then, we would like to estimate the parameters 
of the model, i.e., $a$ and $b$. We will analyze this data with and 
without the hyperparameter method and discuss in detail our results.

\begin{figure}
\begin{center}
\includegraphics[height=5.5cm,width=9cm]{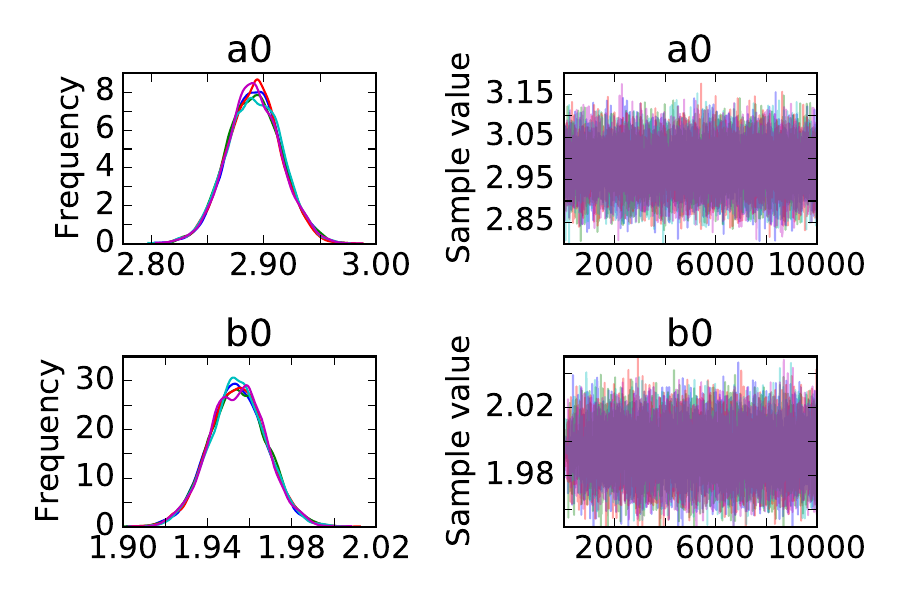} \qquad
\includegraphics[height=4.8cm,width=6.5cm]{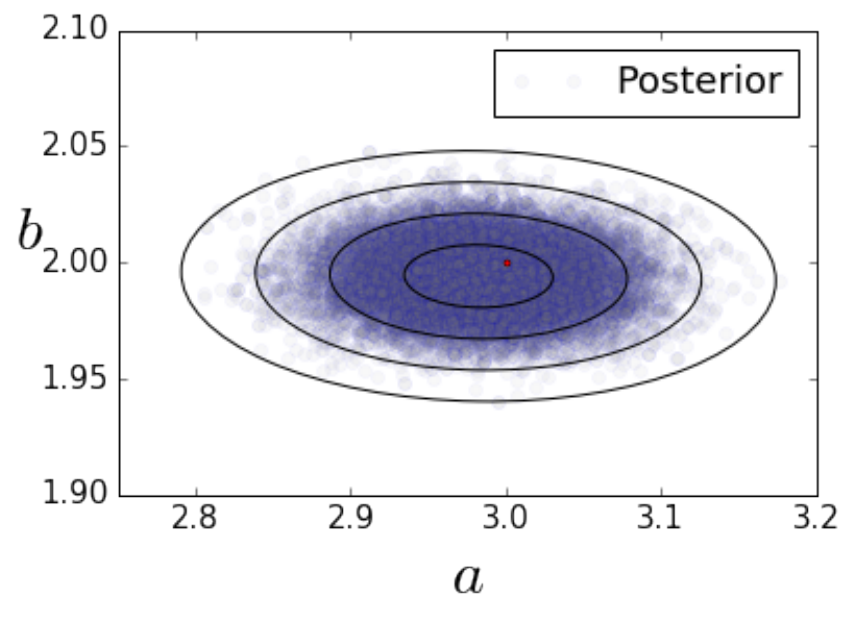}
\end{center}
\caption{\footnotesize{Top panel: 1D marginalized posterior distributions 
for our samples and the Markov chains for model $H_0$. Bottom panel:
2D marginalized posterior distributions along with 1-4 confidence 
regions for our parameters for model $H_0$. The red point corresponds to the 
true value.}}
\label{contour_1}
\end{figure} 

%%%------------------------------------------------------------------------%%%
\textbf{Model $H_0$: {without} hyperparameters.}  \\

%%%....................................................................................%%%
\begin{figure*}
%\begin{floatrow}
%\ffigbox[]
\begin{center}
\includegraphics[height=5.5cm,width=15cm]{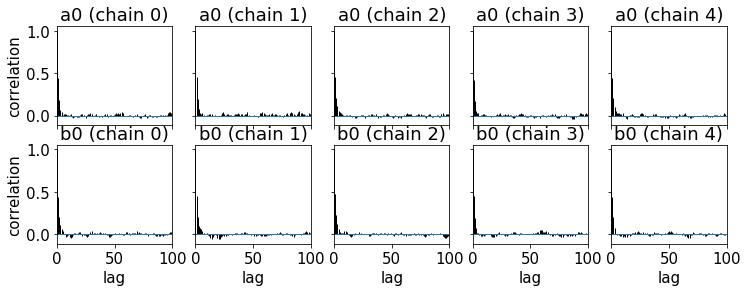}
\end{center}
\caption{\footnotesize{Autocorrelation plots for model $H_0$.}}
\label{autocorr}
%\end{floatrow}
\end{figure*}
 %%%....................................................................................%%%

Before we make a 
Bayesian estimation, it is necessary to specify the priors. As we have 
seen, a good prior is a non-informative one. Suppose we only know some
limits for $a$ and $b$, as can be seen when plotting the data. 
Then, we consider the flat priors%
\begin{equation}
a \propto U[0,5] \ \ \ \text{and} \ \ \ b \propto U[0,3],
\end{equation}
where $U[\alpha,\beta]$ are uniform distributions with lower limit 
$\alpha$ and upper limit $\beta$.

From Equation \eqref{GLik}, we can write our likelihood as%
\begin{equation}
L(D;line)\propto \exp\left[-\sum_d \frac{(y_d-y)^2}{2\sigma_d^2}\right],
\end{equation}
where $y_d$ is our data taken from the dataset $D=D_1+D_2$, and $\sigma_d$ its errors.

We use the MHA to generate the Markov chains. In our analysis, we ran 5 
chains with 10,000 steps for each one. We ran each chain with 
temperature $T=2$ and we thinned them every 50 steps. 
The result corresponds to $a=2.982 \pm 0.047$ and $b=1.994 \pm 0.013$,
and their posterior distributions are plotted in
Figure \ref{contour_1}. Notice that there 
are some regions where the frequency of events in our sample is increased. 
So, we can say that such parameter regions seem more likely to match the data. 
Additionally, we compute the Gelman-Rubin criterion for each variable in order to verify that our results converged, i.e., for $a$ is $1.000017$, 
and for $b$ is $1.000291$. We see that this number is very close to 1, so our 
convergence criterion is achieved.
The bottom panel of Figure \ref{contour_1} displays the $1-4\ \sigma$ confidence 
regions. We also added a point in red to show the real value for our parameters. The real 
value for $a$ and $b$ are within the curve corresponding to one standard 
deviation of our estimations in the inferential method.

As we mentioned, we 
need the autocorrelation to be small as $k$ increases in order to consider that our 
analysis is converging. We see in Figure \ref{autocorr} such plots and 
notice that our convergence criterion is fulfilled.
Then, in Case 1, 
we can see that the model $H_0$ seems to be a very good estimation procedure.

%%%------------------------------------------------------------------------%%%

\textbf{{Model $H_1$: with hyperparameters}.} \\

\begin{figure}[t!]
\begin{center}
\includegraphics[trim = 0mm  0mm 0mm 0mm, clip, width=7.cm, height=4.5cm]{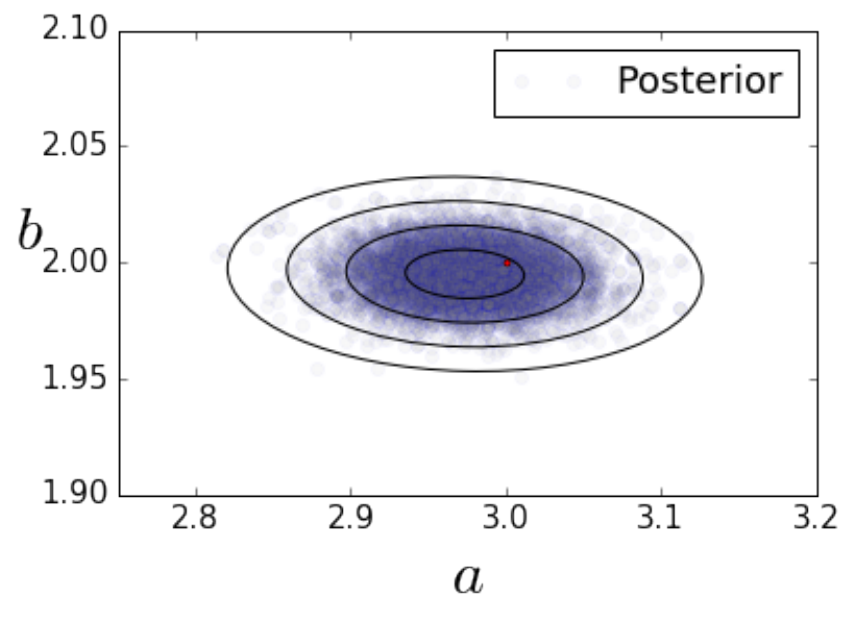} \qquad
\includegraphics[trim = 0mm  0mm 0mm 0mm, clip, width=8.cm, height=4.5cm]{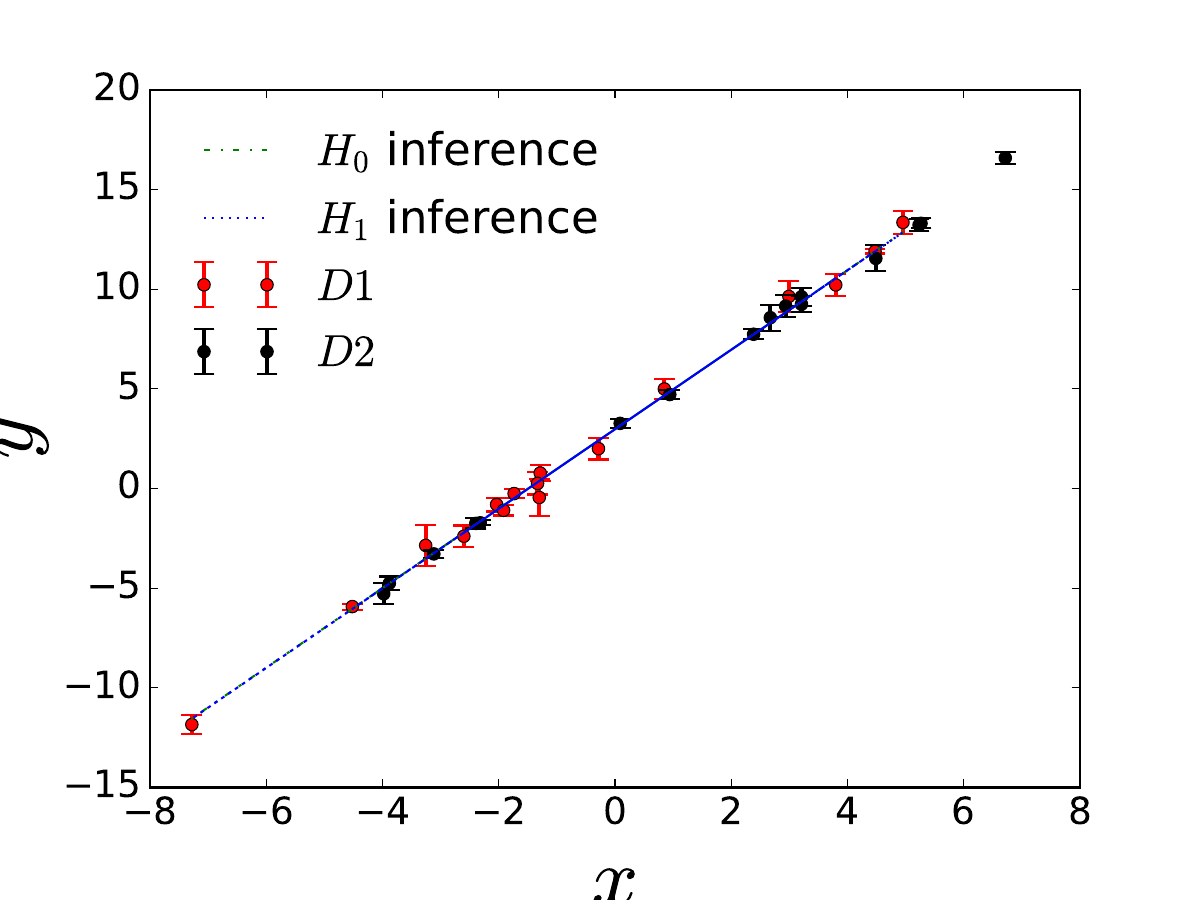}
\end{center}
\caption{\footnotesize{Upper panel: confidence regions for the parameters in model $H_1$.
Lower panel: the best straight-lines inferred, along with  the data.}}
\label{stright1}
\end{figure}

{Now, let us consider the 
Hyperparameter method. In this case, our likelihood can be written 
as Equation \eqref{hyperlik}. Similar to the last procedure, we compute 
the posterior with flat priors, using 5 chains with 10,000 
steps for each one, and check for autocorrelations. 
Our results are as follows: $a=2.97 \pm 0.038$
with Gelman-Rubin of 1.000113, and $b=1.995\pm 0.010$ with Gelman-Rubin 1.000155.
Comparing both procedures, we observe they 
provide similar results. In fact, the confidence regions for both approximations, 
Figure~\ref{contour_1}} and the top panel of Figure \ref{stright1} are similar as well. So, 
which method is better? We could say that the method with hyperparameters 
is as good as the one without them, but, in order to be sure, we compute the evidence ratio $K$ between both models. We obtained 
from Equation~\eqref{bayesfactork} 
\begin{equation}
K = 3.
\end{equation}

{Then, comparing with Table \ref{tab:Jeffrey}, we can say that the evidence 
for $H_1$ to be better than $H_0$ is weak. In such a case, it should 
be equally better to work with $H_0$ as to $H_1$, as explained~before.}

Finally, in order to exemplify our results, let us plot in the bottom panel of
Figure \ref{stright1} our data with the straight-line inferred 
by the mean parameters of both models. As we expected, our 
estimation fits well the data for both cases.

%%%------------------------------------------------------------------------%%%
\subsection{Case 2}
%%%------------------------------------------------------------------------%%%

Here, we consider that we have the same theory for the 
straight-line but different measurements. 
The data points are given in the lower panel of Figure \ref{data_1}. These
correspond to our dataset $D_1$ and $D_2$, but now changing $D_2$ 
by 16 new points generated around the line $y=3.5+1.5x$ with a Gaussian 
noise and standard deviation $\sigma = 0.5$. So, our datasets are not 
auto-consistent with each other. Let us make again the parameter estimation 
for the parameters $a$ and $b$ and look for the differences in both procedures.

\textbf{Model $H_0$: {without} hyperparameters.}  \\

\begin{figure}[h!]
\begin{center}
\includegraphics[trim = 0mm  0mm 0mm 0mm, clip, width=7.cm, height=4.cm]{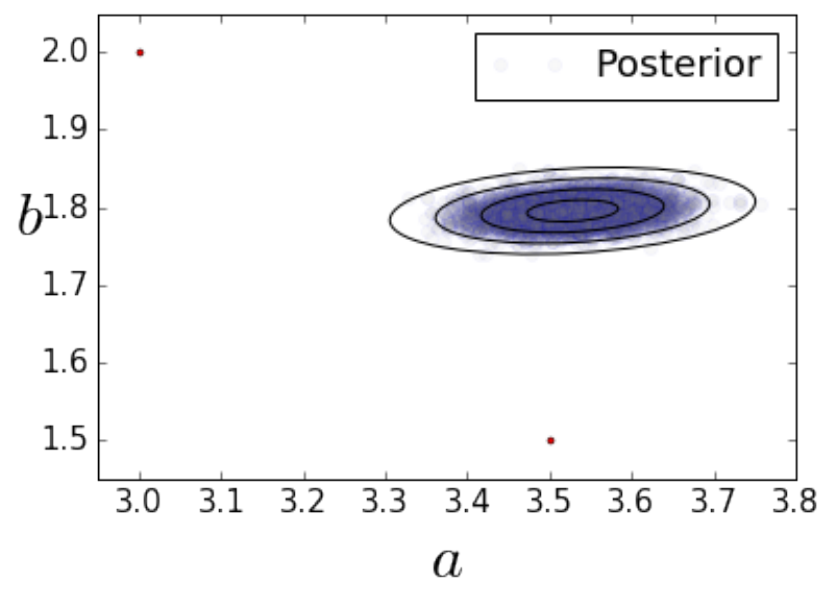} \quad
\includegraphics[trim = 0mm  0mm 0mm 0mm, clip, width=7.cm, height=4.cm]{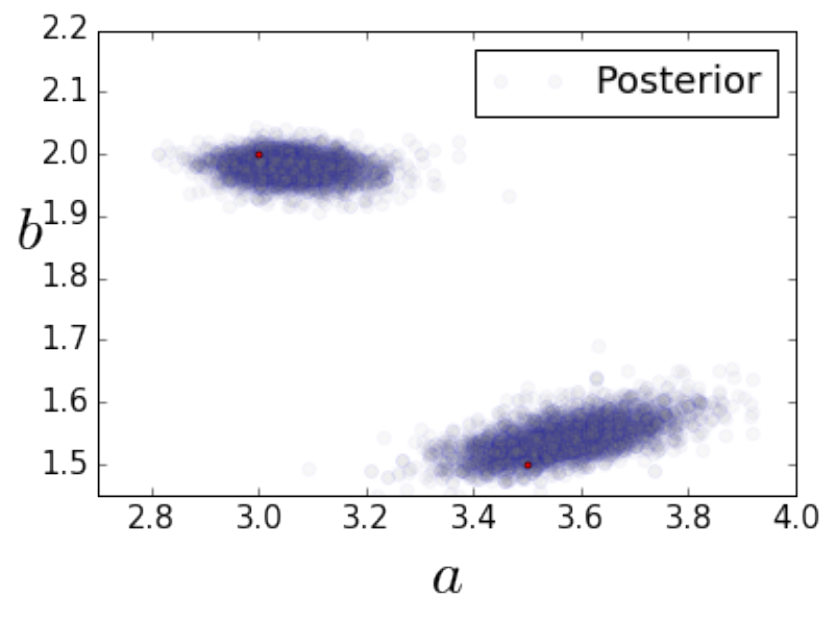} \quad
\includegraphics[trim = 0mm  0mm 0mm 0mm, clip, width=7.cm, height=4.cm]{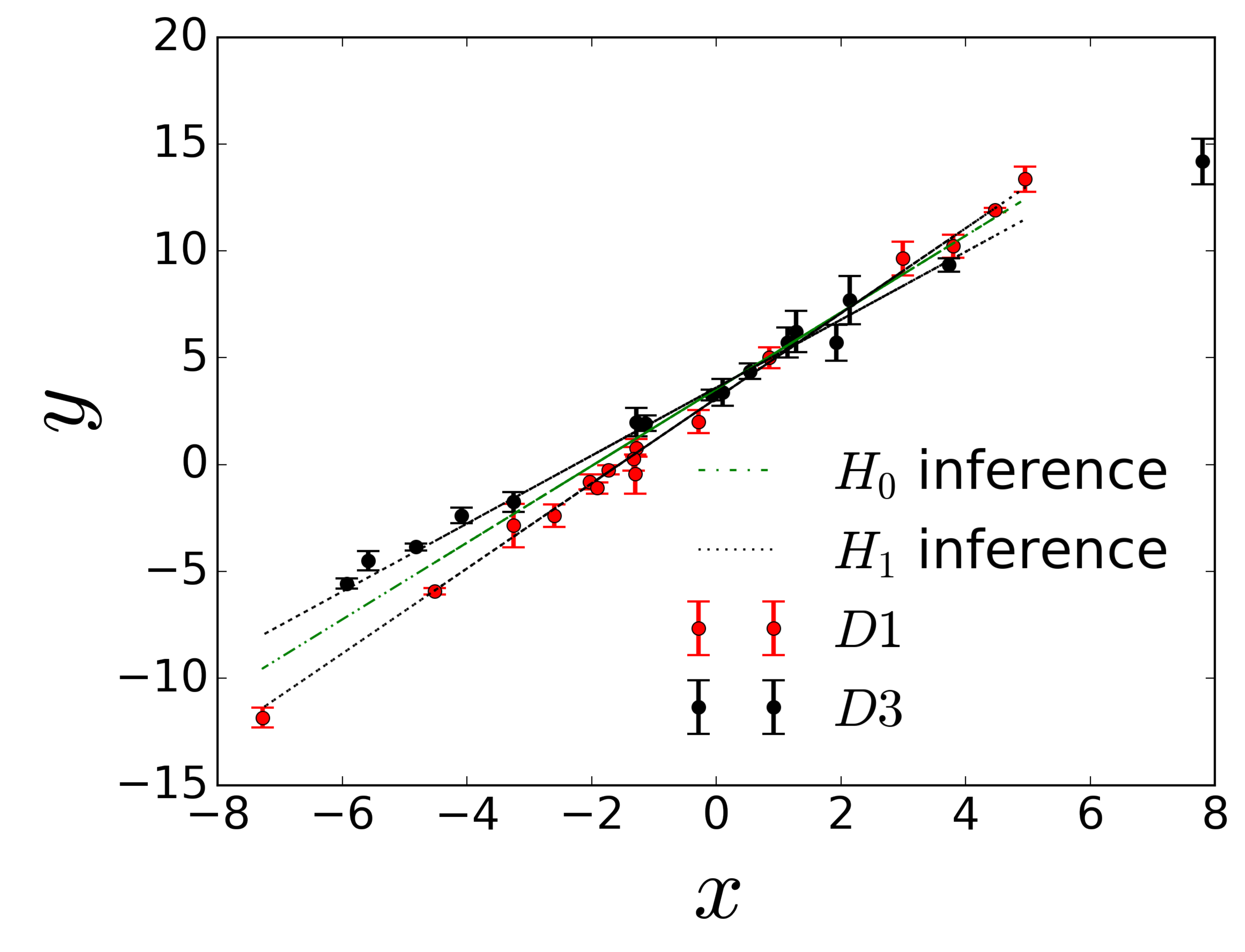}
\end{center}
\caption{\footnotesize{Upper panel: confidence regions for the parameters in model $H_0$.
Middle panel: confidence regions for the parameters in model $H_1$.
Bottom panel: Best-fit values for the straight-lines for Case 2 inferred with the datasets shown.}}
\label{contour_3}
\end{figure}

We follow the same procedure as in Case 1. We computed our posterior and verified that our results converged with the help of the Gelman-Rubin criterion and the autocorrelation plots. Our results are the following: $a=3.528 \pm 0.056$ and $b = 1.795\pm 0.014$.
Then, we plotted $1-4 \sigma$ confidence regions in the upper panel of Figure \ref{contour_3}. 
It is easy to see that our estimation differs so much from the real 
parameters of the datasets (red points). This is because we are trying to fit a model with non-auto-consistent datasets; therefore, we arrive at incorrect results. Now, let us see what happens in the hyperparameters procedure. 

%%%....................................................................................%%%
\textbf{Model $H_1$: {With} hyperparameters.}  \\

In the bottom-left panel of Figure \ref{contour_3}, we plotted the posterior distribution. 
We see immediately that both 
approximations are very different. While, for model $H_0$, we obtained a 
single region far away of the real values of our data, for model $H_1$, 
we obtained two local maximum regions near the real values for our 
datasets (red dots).

Given the fact that we know a priori the real values of the parameters, we could immediately say that the method with 
hyperparameters is a better approximation than the case without them. 
However, we confirm this assumption by computing the ratio $K$ between 
both models. We then obtain%
\begin{equation}
K = 37,
\end{equation}
which means that we have very strong evidence that $H_1$ is better that $H_0$.
Finally, we can plot the straight-line inferred by model $H_0$ and the 
two lines inferred by model $H_1$. Considering the parameters inside the two 
regions in the bottom-left of Figure \ref{contour_3}, we obtain 
the bottom-right panel of the same Figure.

%%%....................................................................................%%%

%%%================================================================%%%
\section{Bayesian statistics in Cosmology}
\label{sec:Cosmology}
%%%================================================================%%%

\subsection{Theoretical Background}

Bayesian statistics is a very useful tool in Cosmology to determine, for instance, the
combination of model parameters that best describes the Universe.
In this section, we present the basics of Cosmology and apply the Bayesian statistics
to perform the parameter inference. In our examples, we will focus on 
the background Universe---and avoid perturbations---since the main purpose of this article is the application of these techniques rather than 
the cosmology by itself.
It should be clear, however, that the extension to 
consider perturbations is immediate, i.e., there is only an increment in the 
number of parameters, and the expressions 
turn out to be just a little more complicated. 

%%%------------------------------------------------------------------------%%%
\subsubsection{Einstein Field equations}
%%%------------------------------------------------------------------------%%%

In order to specify the geometry of the Universe, an essential assumption is the 
\textbf{{Cosmological Principle}}: for a
particular time and on sufficiently large scales, the observable Universe can be 
considered homogeneous
and isotropic, with great precision. For example, at scales greater than 100 
Mega-parsecs, the distribution of galaxies observed on the celestial sphere justifies 
the assumption of isotropy. The uniformity observed in the
temperature distribution (one part in $10^5$) measured through the Cosmic Microwave Background
(CMB) is the best observational evidence we have in favor of a universal isotropy. 
Therefore, if isotropy is taken for granted and by taking into account that our 
position in the Universe has no preference---known as the \textbf{Copernican Principle}---, the homogeneity follows 
when considering isotropy in each
point.

			\textbf{{Homogeneity}} establishes that the Universe is observed equally in each point of space. \par
			\textbf{Isotropy} establishes that the Universe is observed equally in all directions.

The formalism of General Relativity establishes the relationship 
between the geometry of space-time and the matter on it. That is, the 
curvature of the spacetime produces physical effects on the matter, and 
these effects are associated with the gravitational field. Additionally, the 
curvature is related to the matter, described by an energy-momentum tensor 
$T_{\mu \nu}$. 
The above expressions can be summarized by paraphrasing Wheeler: \textit{``matter tells space-time how to curve and, in turn, the geometry of this curvature tells matter how to move''}. 
We can write this sentence down by the Einstein equations
\begin{equation}
\label{einstein}
G_{\mu \nu} = 8\pi G T_{\mu \nu},
\end{equation}
where $G_{\mu \nu}$ is the Einstein tensor (geometry of the spacetime),
and $G$ is the gravitational Newton constant \cite{cambridge, wald, Iorio:2015mga, Debono:2016vkp}.
Throughout this review, we use natural units 
 $c=\hbar=1$.
 
The distance between two points in a curved 
space-time can be measured as
\begin{equation}
ds^2 = g_{\mu \nu}dx^{\mu}dx^{\nu},
\end{equation}
where $g_{\mu \nu}$ is the metric tensor that contains all the information 
about the geometry of the space-time. From now on, and unless stated otherwise, 
greek letters $\mu$, $\nu$, \ldots denote spacetime indices ranging from 0 to 3, 
while latin letters $i$, $j$, \ldots denote spatial coordinates ranging from 1 to 3. 

The geometry that best describes a homogeneous, isotropic, and expanding Universe
is given by the Friedmann-Lema\^itre-Robertson-Walker metric (FLRW), with 
a line element
\begin{equation}
\label{m2}
ds^2 = dt^2 -a^2(t)\gamma_{ij}dx^idx^j ,
\end{equation}
where
\begin{equation}
\label{m3}
\gamma_{ij} \equiv \delta_{ij} + \kappa \frac{x_i x_j}{1 - \kappa \left(x_k x^k\right)}.
\end{equation}

In Equation \eqref{m2}, $a$ represents the scale factor of the Universe 
which only depends on time, and by convention is normalized to today 
$a(t_0) \equiv 1$.
Similarly, in expression \eqref{m3}, $x^i$ labeled the spatial coordinates 
(also called comoving coordinates), $\delta_{ij}$ is the Kronecker delta 
and $\kappa$ describes the curvature of the space-time.

%%%------------------------------------------------------------------------%%%
\subsubsection{Friedmann and continuity equations}
%%%------------------------------------------------------------------------%%%

The content of the Universe needs to satisfy homogeneity and 
isotropy, as well; hence, here it is described by the energy-momentum tensor 
of a perfect fluid 
\begin{equation}
T_{\mu \nu} = (\rho + P)U_{\mu}U_{\nu} - Pg_{\mu \nu},
\end{equation} 
where $\rho$ is the energy density, $P$ is the fluid pressure, and $U_{\mu}$ is the 
4-velocity relative to the observer. If we take the velocity as 
$U^{\mu} = \left(1,0,0,0\right)$ (comoving observer), the energy-momentum tensor reduces to
\begin{equation}
\label{tensorfluidoperfecto}
T^{\mu}_{\nu} = g^{\mu \lambda} T_{\lambda \nu} = \begin{pmatrix}
\rho & 0 & 0 & 0 \\
0 & -P & 0 & 0 \\
0 & 0 & -P & 0 \\
0 & 0 & 0 & -P \\
\end{pmatrix} .
\end{equation}

Using Equations (\ref{einstein}) and (\ref{tensorfluidoperfecto}), with the FLRW metric,
we can obtain the Friedmann~equations
%\begin{subequations}
\begin{equation}
\label{friedmann1}
H^2  \equiv \left(\frac{\dot{a}}{a}\right)^2 = \frac{8 \pi G}{3} \sum_{i}\rho_{i} - \frac{\kappa}{a^2},
\end{equation}
\begin{equation}\label{friedmann1b}
\qquad \qquad \frac{\ddot{a}}{a} = -\frac{4 \pi G}{3} \sum_{i}\left(\rho_i + 3P_i\right).
\end{equation}
%\end{subequations}

In these expressions, $H$ accounts for the rate of expansion/contraction of
the Universe, named as the Hubble parameter.
Subindex $i$ labels all the components that believe the Universe is made of, as we
will see in the next section. 
These equations describe the evolution of the Universe. By combining Equations 
\eqref{friedmann1} and \eqref{friedmann1b}, we can obtain the continuity equation given by
\begin{equation}
\label{continuidad}
\dot{\rho} + 3 \frac{\dot{a}}{a} \left(\rho + P\right) = 0.
\end{equation}

The meaning of \eqref{continuidad} is the conservation of the energy-momentum tensor
($\nabla_\mu T^{\mu \nu}=0$). In order to close the system, we need 
to include an equation-of-state that relates pressure and 
energy density for a given fluid. 
In particular, we are interested on barotropic fluids, which generally
have the form of $P=\omega \rho$.

%%%------------------------------------------------------------------------%%%
\subsubsection{Content of the Universe}
%%%------------------------------------------------------------------------%%%

{Once the equations that define the dynamics of 
the Universe are known, it is necessary to specify its content. The standard cosmological model, also known as $\Lambda$ 
Cold Dark Matter ($\Lambda$CDM), is one of the most accepted models to describe 
the Universe, with its content being: }
\begin{itemize}
	\item \textbf{{Dust:}} It has no pressure, and its energy density takes the 
	form of $\rho \propto a^{-3}.$ Dust is conformed by baryons (ordinary matter).
	
    \item  \textbf{Dark matter:} It is proposed to explain several astrophysical observations, 
    like the dynamics of galaxies in the Coma cluster or the rotation curves of 
    galaxies \cite{liddle, retos}. The $\Lambda$CDM model assumes the dark matter only interacts 
    gravitationally (and possibly by weak interaction) with the rest of the Universe, hence its name, Cold Dark Matter (CDM). Since it is proposed as interacting only via gravitational force, there can be several candidates to fulfill this requirement: it could be conformed by weakly interacting massive particles (WIMPs), by gravitationally-interacting massive particles (GIMPs); by axions (hypothetical elementary particles); or by sterile neutrinos, just to mention a few. 
    For a short review about Dark Matter and possible candidates, see Reference~\cite{arun2017dark}.
    %If the reader wishes to delve further in Dark Matter and some possible candidates in  there is a small review of these and other propositions.
	
	\item \textbf{Radiation:} This corresponds to relativistic particles that follow the relation 
	$P = \frac{1}{3} \rho$, which implies a density with a behavior $\rho \propto a^{-4}$. 
	We consider photons $\rho_{\gamma}$ and massless neutrinos $\rho_{\nu}$
	as radiation, so the total radiation 
	energy density in the Universe is given~by
	\begin{equation}
	    \rho_r = \rho_{\gamma} + \rho_{\nu}.
	\end{equation}
	
	The relation between these quantities is
	\begin{equation}
	    \label{neutrinorelation}
	    \rho_{\nu} = N_{\rm eff}\times\frac{7}{8}\times\left(\frac{4}{11}\right)^{4/3}\rho_{\gamma},
	\end{equation}
	where $N_{\rm eff}$ is the effective number of relativistic degrees of freedom, 
	with standard value $N_{\rm eff} = 3.046$ \cite{parametros}.
	
	\item \textbf{Dark Energy:} It is introduced to explain the current accelerated
	expansion of the Universe. In the $\Lambda$CDM model, dark energy is given 
	by the cosmological constant $\Lambda$ or equivalently 
	by an equation-of-state $\omega=-1$.
\end{itemize}

%%%....................................................................................%%%
\begin{table}[h!]
\begin{center}
\begin{tabular}{cc} 
\cline{1-2}\noalign{\smallskip}
\vspace{0.2cm}
Component & $\omega$\\

\hline
\vspace{0.2cm}
Dust&  0  	 \\
\vspace{0.2cm}
Radiation       &   1/3 \\
\vspace{0.2cm}
Cosmological Constant     &   -1 \\
\hline
\hline
\end{tabular}
\caption{\footnotesize{Equation of state associated to each component of the Universe.}}
\label{state}
\end{center}
\end{table}
%%%....................................................................................%%%

Each of these components can be described by its equation of state
shown in Table \ref{state}, and
defining the density parameter
\begin{equation}
\Omega_{i} \equiv \frac{\rho_{i}}{\rho_{\rm crit}}, \quad {\rm with} \quad 
\rho_{\rm crit} = \frac{3H^2}{8 \pi G},
\end{equation} 
with $\rho_{\rm crit}$ being the condition to have a 
flat Universe or equivalently zero curvature, we can rewrite (\ref{friedmann1}) as
\begin{equation}
\label{friedmann3}
\frac{H^2}{H_0^2} = \Omega_{r,0} a^{-4} + \Omega_{m,0} a^{-3} + \Omega_{k,0} a^{-2} + \Omega_{\Lambda,0},
\end{equation}
where $\Omega_{r,0}$ is the radiation density parameter, 
$\Omega_{m,0}\equiv \Omega_{b,0}+\Omega_{DM,0}$ corresponds to the total matter, 
$\Omega_{b,0}$ to baryons, $\Omega_{DM,0}$ to dark matter, $\Omega_k \equiv -\kappa/(a H)^2$ the curvature 
density parameter, 
and $\Omega_{\Lambda} \equiv \Lambda /3H^2$ associated with the Cosmological Constant, and the 
subscript zero indicates they are evaluated by %Please ensure the meaning has been retained.
today's values.

%%%------------------------------------------------------------------------%%%
\subsubsection{Alternatives to the $\Lambda$CDM model} \label{alternatives}
%%%------------------------------------------------------------------------%%%

{The $\Lambda$CDM model has had great success in modeling a wide range of astronomical
observations. However, it is in apparent conflict with some observations on 
small-scales within galaxies (e.g., cuspy halo density profiles, overproduction 
of satellite dwarfs within the Local Group, amongst many others; see, for example, Reference
\cite{retos,beyondCDM}). 
In addition, all attempts to detect WIMPs either directly in the laboratory,
or indirectly by astronomical signals of distant objects have failed 
so far. For some of these reasons, it seems necessary to explore alternatives to the 
standard $\Lambda$CDM model. 
With this in mind, several alternatives have been suggested. 
For instance, the Scalar Field Dark Matter (SFDM) model proposes the 
dark matter as a spin 0 boson particle \cite{Matos:2009hf, ideas1, ideas2, ideas3, ideas4, Matos09};
or the 
Self Interacting Dark Matter, as its name states, relies on the cold dark 
matter to be made of self interacting particles \cite{sidm, Gonzalez:2008wa, Padilla:2019fju}.
On the other hand, in order to explain the accelerated expansion of the Universe,
there exist different 
modifications to the theory of General Relativity, i.e., $f(R)$ theories \cite{Felice100}, 
braneworld models \cite{modifiedgravity, Miguel18}.
There are also some alternatives to the cosmological constant as Dark Energy, 
i.e., scalar fields (quintessence, K-essence, phantom, quintom, 
non-minimally coupled scalar fields
\cite{Vazquez:2020ani, Akarsu:2020pka, quintessence, modelsDE, quintom}); or many more alternatives, i.e.,
anisotropic Universes \cite{Akarsu19,anisotropic1,anisotropic2}.}
Finally, if the dark energy is assumed to be a perfect fluid, one of the most popular time-evolving parameterizations for its equation of state
consists of expanding $\omega$ in a Taylor series, 
for example, the Chevallier-Polarski-Linder (CPL) 
$\omega = \omega_0 + \omega_a\left(1-a\right)$, with
two free parameters $\omega_0, \omega_a$ \cite{CPL1, CPL2}. It may also be 
expanded into Fourier series \cite{fourier}, or many more Bayesian approaches, as
have been suggested to account for a dynamical dark energy \cite{Vazquez:2012ce, Hee16, Vazquez:2012ag}.

%%%------------------------------------------------------------------------%%%
\subsection{Cosmological parameters}
%%%------------------------------------------------------------------------%%%

%%%------------------------------------------------------------------------
\subsubsection{Base parameters}
%%%------------------------------------------------------------------------
These parameters, also known as \textit{standard parameters}, are the main quantities 
used in the description of the Universe. They are not predicted by a fundamental 
theory, but their values must be fitted to provide the best description 
of the current astrophysical and cosmological observables. 
To explain the homogeneous and isotropic Universe, we can use the density 
parameter of each component $\Omega_{i,0}$ and the Hubble parameter $H_0$ related 
by~\eqref{friedmann3}. 
In particular, the radiation contribution is measured with great precision, so that
$\Omega_\gamma$ is pinned down very accurately, and, hence, there is no need to fit this parameter. 
Similarly, %Please ensure the meaning has been retained. %Authors: we confirm that.
neutrinos, as long as they maintain a relativistic behavior, can be related to the density of the photons through \eqref{neutrinorelation}. 

On the other hand, the existence of strong degeneracies from different combinations 
of parameters is also notorious. In particular, the geometric degeneracy involving 
$\Omega_m$, $\Omega_{\Lambda}$ and the curvature parameter 
$\Omega_k = 1-\Omega_m - \Omega_{\Lambda}$. To reduce these degeneracies, 
it is common to introduce a combination of cosmological parameters such that 
they have orthogonal effects in the measurements.

%%%------------------------------------------------------------------------%%%
\subsubsection{Derived parameters}
%%%------------------------------------------------------------------------%%%

The above standard set of parameters provides an adequate description of the 
cosmological models. However, this parameterization is not unique, and some 
others can be as good as this one. Various parameterizations make use of the 
knowledge of the physics or the sensitivity of the detectors and can, 
therefore, be interpreted more naturally. In general, other parameters 
could have been used to describe the Universe, for example: the age of the 
Universe, the current temperature of the neutrino background, the epoch of 
equality of matter-radiation, or the epoch of reionization. In the standard 
cosmological model ($\Lambda$CDM), in order to decrease degeneracies, the 
physical energy densities $\Omega_{DM,0}h^2$ and $\Omega_{b,0}h^2$ are used as 
base parameters \cite{parametros}.

%%%================================================================%%%
\subsection{Cosmological observations}
\label{sec:observations}
%%%================================================================%%%

In this section, we review some of the most 
common experiments and observables used to constrain the cosmological models
on the background level.

\textbf{{Baryon} Acoustic Oscillations (BAO):}
The BAO is a statistical property, a feature in the correlation function of galaxies or in the power spectrum. The best description of the early Universe considers that it was made of plasma of coupled photons and matter (baryons and dark matter). The interaction between the gravitational force due to matter and the radiation pressure formed spherical waves in the plasma. When the Universe cooled down enough, the protons and electrons were able to join together, forming hydrogen atoms; therefore, this process allowed photons to decouple from the rest of the baryons. The photons began to travel uninterrupted, while the gravitational field attracted matter towards the center of the spherical wave. The final configuration is an overdensity of matter in the center and a shell of baryons of fixed radius called sound horizon. This radius, used as a standard ruler, is the maximum distance that sound waves could have traveled through the primordial plasma before recombination. The sound horizon $r_d$ is given by
\begin{equation}
    r_d = \int_{z_d}^{\infty} \frac{c_s(z)}{H(z)}dz,
    \label{soundhorizon}
\end{equation}
where the sound speed (in terms of redshift $z$) in the photon-baryon fluid is $c_s(z) = 3^{-1/2}c\left[1+\frac{3}{4}\rho_b(z)/\rho_{\gamma}(z)\right]^{-1/2}$, and $z_d$ is the redshift when photons and baryons decouple.

The BAO scale is determined by adopting a fiducial model to be able to translate the angular and redshift separations at comoving distances. The information of the measurement is found in the ratio ($\alpha$) of the measured BAO scale and that predicted by the fiducial model ($fid$). In an anisotropic fit, two ratios are used, one perpendicular $\alpha_{\bot}$ and one parallel $\alpha_{\parallel}$ to the line of sight. A measurement of $\alpha_{\bot}$ constrains the ratio of the comoving angular diameter distance to the sound horizon \cite{BAOimplications}:
\begin{equation}
    \frac{D_M(z)}{r_d} = \alpha_{\bot}\frac{D_{M,fid}(z)}{r_{d,fid}},
    \label{DM}
\end{equation}
where the comoving angular diameter distance is given by
\begin{equation}\label{D_M}
    D_M(z) = \frac{c}{H_0}S_k\left(\frac{D_c(z)}{c/H_0}\right).
\end{equation}

The line-of-sight comoving distance is defined as
\begin{equation}
    D_c(z) = \frac{c}{H_0}\int_0^z dz'\frac{H_0}{H(z')},
\end{equation}
and $S_k (z)$
\begin{equation}
S_k(x) = 
\begin{cases}
\sinh \left(\sqrt{\Omega_k}x\right)/\sqrt{\Omega_k} & \Omega_k > 0,\\
x & \Omega_k=0,\\
\sin \left(\sqrt{-\Omega_k}x\right)/\sqrt{-\Omega_k}  & \Omega_k < 0.
\end{cases}
\end{equation}

{The Hubble parameter can be constrained by measuring $\alpha_{\parallel}$ using an analogous~quantity}
\begin{equation}
    \frac{D_H(z)}{r_d} = \alpha_{\parallel} \frac{D_{H,fid}(z)}{r_{d,fid}},
    \label{DH}
\end{equation}
with $D_H(z) = c/H(z)$.
\noindent

If redshift-space distortions are weak (which is valid for luminous galaxy surveys but not for the Ly-$\alpha$), an isotropic analysis measures an effective 
combination of \eqref{DM} and \eqref{DH}, 
and the volume averaged distance $D_V(z)$ \cite{BAOimplications}
\begin{equation}
    \frac{D_V(z)}{r_d} = \alpha \frac{D_{V,fid}(z)}{r_{d,fid}},
    \label{DV}
\end{equation}
with $D_V(z) = \left[zD_H(z)D_M^2(z)\right]^{1/3}$.

\begin{figure}[t]
	\begin{center}
	\includegraphics[height=5cm, width=8.5cm]{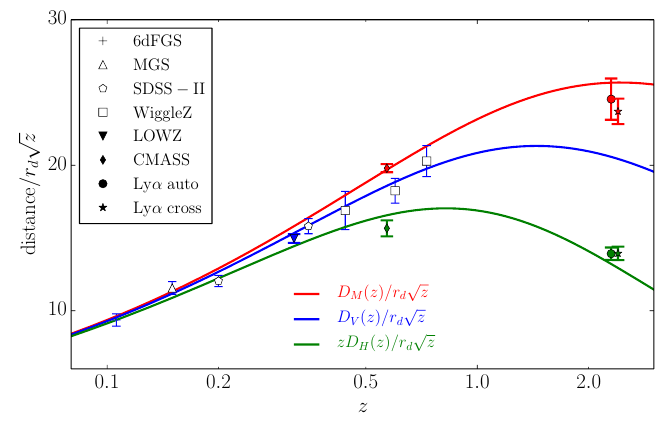}
	\end{center}
	\caption{\footnotesize{BAO Hubble diagram.  BAO measurements of 
	$D_V/r_d$, $D_M/r_d$ and $zD_H/r_d$ from the sources indicated in the legend.
	The scaling factor $\sqrt{z}$ is included for a better display of the 
	error bars. Solid lines are plotted by using the best-fit values obtained 
	by the Planck satellite \cite{Planckcollaboration}. The Ly$\alpha$ cross-correlation points have been shifted in redshift; auto-correlation points are plotted at the correct effective redshift.}}
	\label{baohubble}
\end{figure}

The BAO measurements constrain the cosmological parameters through the radius of 
the sound horizon $r_d$, Hubble distance $D_H(z)$ and the comoving angular diameter 
distance $D_M(z)$; see Figure \ref{baohubble}.
The data used in this work to constrain $D_V / r_d$ is obtained from the 6dF 
Galaxy Survey (6dFGS \cite{6dFGS}) from UK Schmidt Telescope, the Main Galaxy Sample 
(MGS \cite{MGS}), and the 
BOSS LOWZ Sample \cite{BOSSgalaxies} from SDSS. On the other hand, the data from 
BOSS CMASS Sample \cite{BOSSgalaxies}, BOSS 
Lyman $\alpha$ auto-correlation (Ly$\alpha$ auto,~\cite{Lyafauto}), and BOSS Lyman-$\alpha$ cross 
correlation (Ly$\alpha$ cross \cite{Lyafcross}) are used to constrain 
$D_M/r_d$, $D_H/r_d$.

%%%....................................................................................%%%
%%%....................................................................................%%%

\textbf{{Supernovae} type Ia (SNIa):} This type of supernova occurs in binary star systems, one of which is a white dwarf that accretes matter from the star that accompanies it. When the white dwarf accumulates sufficient mass ($\approx$1.4 solar masses), its core will start the ignition temperature for carbon fusion, and, within a few seconds, it releases enough energy to produce the supernova \cite{supernovadef}. 
Since type Ia supernovae (SNIa) are hypothesized to occur near the same mass limit of 1.4 solar masses, commonly referred to as Chandrasekhar mass, their luminosity peaks are fairly consistent and can be standardized and, thus, be used as standard candles \cite{parametros}. From several analyses of SNIa, the Supernova 
Cosmology Project and High-z Supernova Search Team both found evidence that the 
Universe is currently expanding at an accelerated rate \cite{supernova1,supernova2,supernova3}.
 
These stars allow us to measure relative distances  
using the luminosity distance given by
\begin{equation}
    D_L \equiv \sqrt{\frac{L}{4\pi S}},
\end{equation}
where $L$ is the luminosity defined as the energy emitted per unit solid angle 
per second, and $S$ is the radiation flux density defined as the energy received 
per unit area per second \cite{liddle}. The observable quantity is the radiation 
flux density received, and 
it cannot be translated into the luminosity density unless the absolute luminosity 
of the object is known. Even if the luminosity is unknown, 
it will appear as a scaling factor \cite{liddle}. The relation between $D_L$ and the
cosmological parameters is given by
\begin{equation}
    D_L = D_M (1+z),
    \label{DL}
\end{equation}
where $D_M$ is provided by Equation \eqref{D_M}. Another important quantity
in the observation of supernovae is the standardized distance modulus 
\begin{equation}
    \mu = m_B^* - M_B + \alpha X_1 -\beta C,
\end{equation}
where $m_B^*$ is the observed peak magnitude in the rest frame of blue band (B), 
$\alpha$, $\beta$, and $M_B$ are parameters that depend on 
host galaxy properties \cite{jla}. $X_1$ is the
time stretching of the light curve, and $C$ is the supernova color at maximum 
brightness. The relation between the standardized distance modulus and the 
luminosity distance is 
\begin{equation}
    \mu = 5 log_{10}\left(\frac{D_L}{10pc}\right).
\end{equation}

\begin{figure}[t]
	\centering
	\includegraphics[height=5cm,width=8cm]{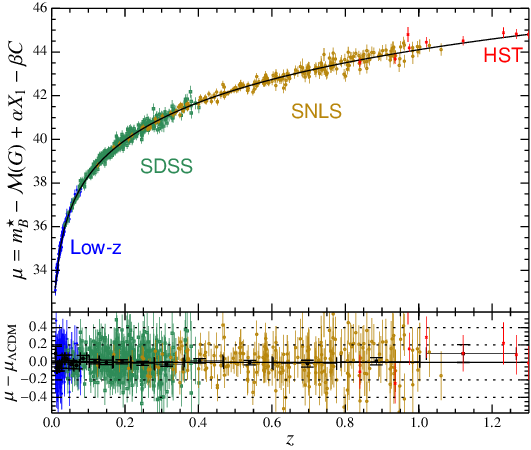}	
	\caption{\footnotesize{Joint Light-curve Analysis data (JLA). Vertical axis is the standardized distance modulus $\mu$ (luminosity distance function) and the horizontal axis is the redshift $z$. Source: \cite{jla2}.}}
	\label{jladata}
\end{figure}

The data used in this paper (shown in Figure \ref{jladata}) is obtained from 
the Joint Light-curve Analysis (JLA). It is a collaboration to analyze the data 
of 740 stars from the SDSS-II (previous version of BOSS), the SuperNova Legacy Survey (SNLS \cite{snls}) experiment that used the Canada-France-Hawaii telescope (CFHT), 
the Cal\'an/Totolo Survey, the Carnegie Supernova Project, the 
Harvard-Smithsonian Center for Astrophysics, La Silla Observatory, 
Fred Lawrence Observatory, and the Hubble Space Telescope (HST) \cite{jla}. 
For simplicity, we compress all the information into a linear 
function fit over 30 bins (31 nodes), spaced evenly 
in $log(z)$ with a $31\times 31$ covariance matrix \cite{BAOimplications}.

Although SNIa have been widely used to restrict cosmological models, there are still discussions about the way it is done. That is, in order to extract information from them, an a priori 
cosmological model has to be assumed, which may biased the outcomes \cite{universe4060073}.

%%%%....................................................................................%%%
%%%....................................................................................%%%

\textbf{{Cosmic }Microwave Background (CMB):} Corresponds to 
the radiation that permeates all the Universe, discovered in 1965. 
Before recombination, baryons and 
photons were tightly coupled, and once photons decouple from the rest of the matter, 
they traveled uninterrupted until reach us. 
The temperature radiation measured at different parts of the sky contains
information of the last scattering epoch, gravitational lensing, among others.
Here, the CMB displays the primordial anisotropies studied in the 
angular power spectrum.
One of the most important recent collaborations that studies the CMB corresponds to the Planck
satellite, and previous probes include COBE \cite{COBE} and WMAP \cite{WMAP}.
It is a European Space Agency mission, in which the main objective is to measure 
the temperature, polarization, and anisotropies of the CMB over the entire sky. 
These results would allow to determine the properties of the Universe at large scales and
the nature of dark matter and dark energy, as well as to test inflationary theories, 
determine whether the Universe is homogeneous or not, and obtain maps of galaxies 
in the microwave \cite{Planck1, Planck2, Planck3,cmb1,cmb2,cmb3,cmb4,cmb5}.

In this work, we use the CMB information as a BAO located at redshift $z=1090$, 
measuring the angular scale of the sound horizon $D_M(1090)/r_d$, in addition, to calibrating the absolute length of the BAO ruler through the 
determination of $\Omega_b h^2$ and $\Omega_{cb} h^2$ (the density parameters of baryonic and dark matter, respectively; more details about these parameters can be found in Reference \cite{BAOimplications}).

\textbf{{Large} Red Galaxies:} (Cosmic Chronometers) These are the most massive galaxies for each redshift $z$ and contain the oldest star population. These kinds of galaxies are used to estimate the Hubble factor because they contain little stardust, which makes it easier to get their spectra. The way these chronometers work is by selecting two galaxies at different redshifts between $z \sim 0-2$ and compare their upper cut in its age distributions. By doing this, it is possible to obtain the difference of ages $\Delta t$ and redshifts $\Delta z$ such that the expression $\frac{dz}{dt}$ can be approximated as $\frac{dz}{dt}\simeq\frac{\Delta z}{\Delta t}$. This quantity is related with the Hubble factor via \cite{H1, H2}
\begin{equation}
H(z) = \frac{-1}{1+z} \frac{dz}{dt}\simeq\frac{-1}{1+z}\frac{\Delta z}{\Delta t}.
\label{HZ}
\end{equation}

{In this work, the data used to constrain $H(z)$ was obtained from Reference \cite{H3,H4,H5, H6}. 
In Figure~\ref{HzData}}, a compilation of these data is shown. 

Another important probe in the foreseen future corresponds to the Gravitational waves from merging black holes (standard sirens), as they directly measure the luminosity distance to the merger. 
 Therefore, it is possible to constrain the distance-redshift relation and, hence, the cosmological parameters \cite{Abbott:2017xzu,Guidorzi:2017ogy}.
 
Even though this paper is focused on cosmological constraints on the background level, there exist several observations to pin down the parameters at the perturbation level, and some of them include 
gravitational lensing \citep{lensing1,lensing2,lensing3,lensing4,lensing5}, clustering \citep{cluster1,cluster2,cluster3,cluster4}, the matter power spectrum \citep{mps1,mps2}, and primordial gravitational waves and the primordial power spectrum, both generated during inflation \citep{inf1,inf2,inf3}, among many others. 
%

 %%%================================================================%%%
\section{Parameters inference in Cosmology}
\label{example1}
%%%================================================================%%%

%%%%....................................................................................%%%
\begin{figure}
	%\begin{center}
\includegraphics[trim = 0mm  0mm 0mm 0mm, clip, width=7.5cm, height=5cm]{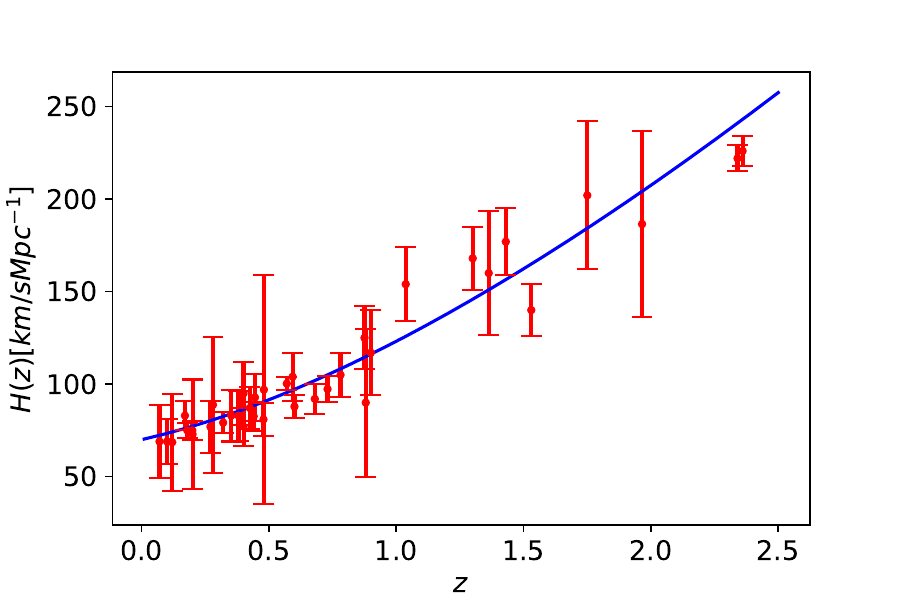}\qquad
	\caption{\footnotesize{Hubble data as a function of redshift $z$ for Cosmic Chronometers. The solid line corresponds to the best-fit by using $\Lambda$CDM model.}}
	\label{HzData}
	%\end{center}
\end{figure}
%%%....................................................................................%%%

%%%....................................................................................%%%
\begin{figure}[t]
\includegraphics[trim = 0mm  0mm 0mm 0mm, clip, width=8.cm, height=4.5cm]{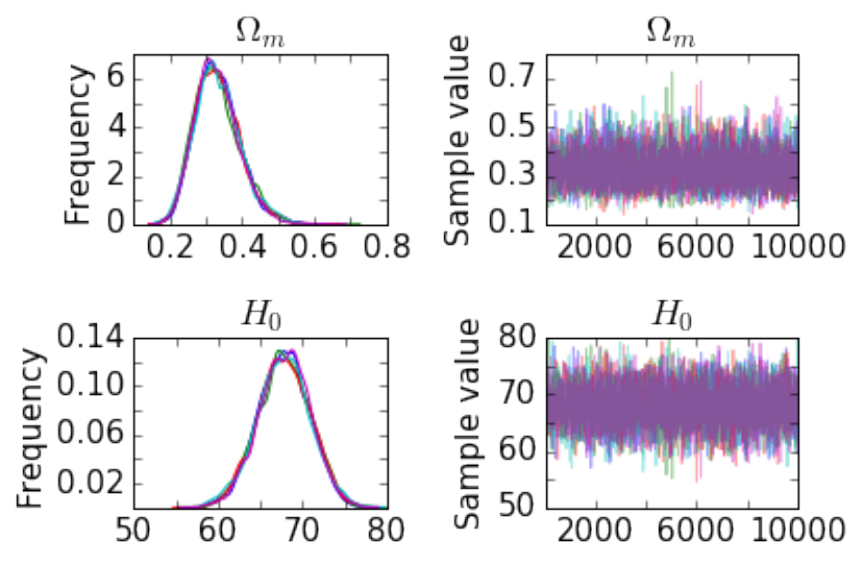}	
	\includegraphics[trim = 0mm  0mm 0mm 0mm, clip, width=8.cm, height=4.5cm]{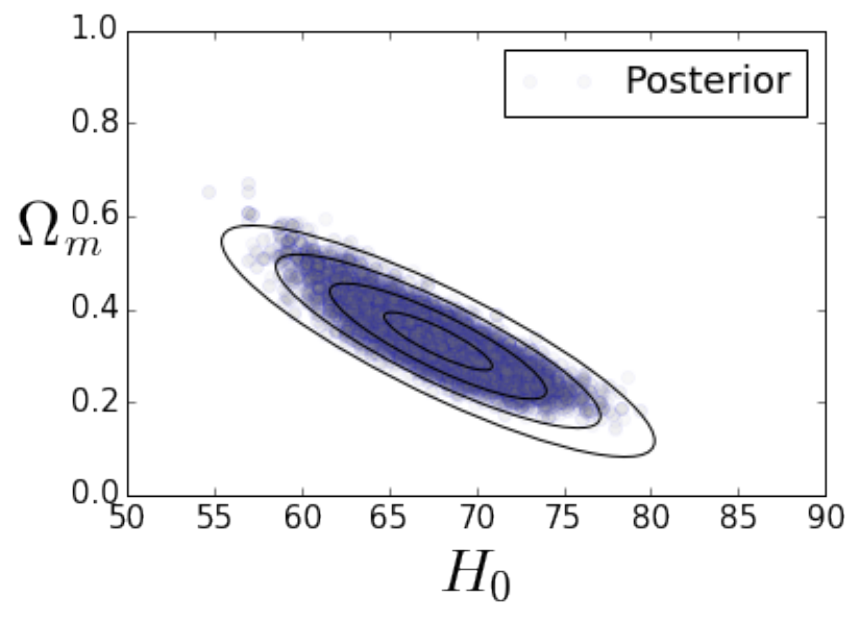}
\caption{\footnotesize{Results for the $\Lambda$CDM model.
    Upper panel: 1D posterior distributions for the parameters $H_0$ and 
    $\Omega_m$ along with its chains. Lower panel: joint 2D posterior 
    distributions with 1-4 confidence levels.}}
	\label{chains_lcdm}
\end{figure}
%%%....................................................................................%%%

Here, we estimate the Hubble parameter $H_0$ at the present time and the matter density 
 of the Universe $\Omega_m$, assuming a $\Lambda$CDM standard model. 
We use our own Python code, which can be found in 
Reference \cite{git}. 

\textbf{{The data.-}}
For this particular example, we focus only on the Cosmic Chronometers, 
as shown in Figure \ref{HzData}.
%MDPI: Please confirm whether the italics are necessary. 
%Authors: We change them for bold. They're not necessary, but we'd like to stress out these concepts(The data, The model and Numerical estimation).

\textbf{{The model.-}} Our interest is to fit the density parameters 
for each component of the Universe, as well as the value of the Hubble parameter 
at present time. 
For this purpose, we use Equation \eqref{friedmann3}, 
and, for simplicity, we assume a flat Universe with a well measured radiation content.
Then, the model is given by
\begin{equation}\label{H2}
H^2 = H_0^2[\Omega_{m,0} (1+z)^{3}+ \Omega_{\Lambda,0}],
\end{equation}
constrained by the relation
\begin{equation}
\Omega_{m,0}+\Omega_{\Lambda,0}=1.
\end{equation}

Notice that the above relation implies that we can get rid of a parameter---$\Omega_{m,0}$ or $\Omega_{\Lambda,0}$---in the analysis. Then, the parameters 
we decided to estimate are $H_0$ and $\Omega_{m,0}$. By assuming the Gaussian 
approximation, we construct the likelihood given by
\begin{equation}\label{liklcdm}
    L \propto \exp\left[-\sum_i\frac{(H_i-H(z_i))^2}{2\sigma_i^2}\right],
\end{equation}
where $H(z_i)$ is described by Equation \eqref{H2} evaluated at each redshift $z_i$, and
$H_i$ is the value of the Hubble parameter measured at $z_i$ and $\sigma_i$ 
the error of the $i$-th measurement. 

%Now, as part of the model construction, it is necessary to specify our priors. 
The only a priori information we have about the free parameters is that
each component of the Universe $\Omega_{i,0}$ must satisfy the 
relation $0\leq \Omega_{i,0}\leq 1$, while, for the present Hubble parameter,
its conservative prior can be obtained by observing the data at our disposal.
In such cases, a good prior choice is a uniform distribution (flat prior)
with limits $\Omega_m\in [0,1]$ and $H_0\in [10,100]$.
Hence, the priors
\begin{equation}\label{priorlcdmf  }
\Omega_{m,0}\sim U[0,1],\qquad  H_0\sim U[10,100].
\end{equation}

\textbf{{Numerical estimation}}.- We follow the same procedure as in the straight-line
example. In the left panel of Figure \ref{chains_lcdm}, we have plotted the chains
obtained for our estimations and its corresponding 1D posterior distribution. 
Similar to the above examples, we have also plotted 
the 2D posterior distributions with $1$--$4\sigma$ confidence regions in the right panel of the same figure. 
Additionally, we have obtained the mean, standard deviation and the Gelman Rubin
criteria for each parameter: $H_0 =67.77 \pm 3.13$ with convergence 1.00045, and 
$\Omega_m = 0.331 \pm 0.0628$ with convergence 1.00044. 
We can see that our estimations are very similar to
the values reported in the literature \cite{Planckcollaboration}.

%%%================================================================%%%
\subsection{Cosmological and statistical codes}
%%%================================================================%%%

As we have shown, until now, during the process of parameter inference 
for a given model, there are three steps we have followed: 
first, obtain the data we would like to confront with the model parameters. 
Then, construct the likelihood associated with the theory 
we are working with. Depending on the nature of the data, 
the likelihood can depend on different ways of the parameters. 
For example, in the last exercise, the likelihood in terms of the parameters is 
via \eqref{H2} and \eqref{liklcdm}. Finally, it is necessary to program 
the numerical tools in order to obtain the parameter inference. 
Such programming can be done, for example, in PyMC3, as we saw before. 

It is notorious that the above process can be a tremendous task for 
programming, for instance, when the theories involve numerous parameters,
or some of them must be marginalized in order to ignore the non-interesting ones 
(like nuisance parameters), or/and when the models depend 
on the parameters of interest in a difficult way (like by solving 
differential equations, integrals, etc.). Then, we can proceed in two 
different ways: the first is by accepting the challenge and creating our own code. 
Developing a new code may be a good option 
rather than using others, as it would require a long time to learn 
the implementation and modification (specially if the theory 
we are dealing with is somehow simple).
Otherwise, we need to rely on existing codes, principally when the theory of interest 
is quite complicated, as it is usually the case when perturbations are taken into account
on the cosmological models. 
In this section, we present 
some cosmological and statistical codes to test cosmological models.

%%%================================================================%%%
\subsubsection{Cosmological codes}
%%%================================================================%%%

Today, there are several cosmological Boltzmann codes available. Some of them include: 
CMBFAST (written in FORTRAN 77 \cite{Seljak:1996is, Zaldarriaga:1997va, Zaldarriaga:1999ep}), CMBEASY (C++ \cite{Doran:2003sy}), 
CAMB (FORTRAN 90 \cite{Lewis:1999bs, Howlett:2012mh}), CLASS (C \cite{class1, mont1}), and COSMOSIS \cite{cosmosis} (written in Python, and it works as an 
interface between CLASS, CAMB, MontePython, CosmoMC, and more). All of them are used 
for calculating the linear CMB anisotropy spectra, based on integrations over the 
sources along the photon line of sight.

%%%================================================================%%%
\subsubsection{Statistical codes}
%%%================================================================%%%

Once the cosmological model is established, we need a statistical code 
to estimate the free parameters of our model. There 
are several MCMC codes that can make this task 
easy to handle. Some of them are:

\textit{{Monte Python}} \cite{Brinckmann:2018cvx, Audren:2012wb, MP2}.-
It is a Monte Carlo code for Cosmological Parameter extraction that contains likelihoods of most recent experiments, 
and interfaces with the Boltzmann code Class for computing the cosmological 
observables. The code has several sampling methods available: 
Metropolis-Hastings, Nested Sampling (through MultiNest), 
EMCEE (through CosmoHammer), and Importance Sampling.

\textit{CosmoMC} \cite{Lewis:2002ah}.- This is a Fortran 
MCMC engine for exploring cosmological parameter space. 
It contains Monte Carlo samples and importance sampling. 
It also has by default several likelihoods of the most 
recent experiments, and interfaces with CAMB.

\textit{SimpleMC}.- It is an MCMC code for cosmological parameter 
estimation where only the expansion history of the Universe matters. 
It was written by An\^ze Slosar and 
Jos\'e A. V\'azquez, initially released in Reference \cite{BAOimplications}, and can be downloaded in Reference \cite{simplemc}.
This code solves the cosmological equations for the background parameters 
in the same way as CLASS or CAMB, and it contains the statistical 
parameter inference of CosmoMC/MontePython. An advantage 
of this code is that it is completely written in Python
with an interface to machine learning tools,
such as artificial neural networks, genetic algorithms, as well as algorithms to 
compute the Bayesian evidence, i.e., Dynesty \cite{speagle2020dynesty} or MCEvidence \cite{MCEvidence}. 

The main idea of MCEvidence is that, asymptotically, the number density $n$ of the MCMC is proportional to the density of the likelihood multiplied by the prior, that is, the non-normalized posterior
\begin{equation}
    \Tilde{P}(D|\theta,H) = P(D|\theta,H)P(\theta|H) = an(D|\theta,H),  \nonumber
\end{equation}
where $n(D|\theta,H) = N P(D|\theta,H)$ with $N$ the length of the chain. The code uses Bayesian inference to find $a$, the constant of proportionality (see Reference \cite{Heavens:2017afc} for details). Once this constant has been found, the Bayesian evidence is given by
$P(D|H) = aN.$

%%%================================================================%%%
\section{Examples with SimpleMC}
\label{sec:Simplemc}
%%%================================================================%%%

The main interest of this section is to test several cosmological models
through SimpleMC.
Even though we selected this code, the results we present here can also 
be obtained in any of the aforementioned codes. 

Throughout these examples, we consider Gaussian likelihoods 
for each dataset with the following form:
\begin{equation}
L\propto \exp\left[-\sum_i \frac{(T(z_i)-T_i)^2}{2\sigma_i}\right],
\end{equation}
where $T(z_i)$ is the theoretical value related to 
the observation $T_i$; and $\sigma_i$ are the corresponding errors associated 
to each measurement. In our estimation, $T(z_i)$ is given by \eqref{DL} for Supernovae; \eqref{DM} for CMB; \eqref{DM}, \eqref{DH}, and \eqref{DV} for BAO; and \eqref{HZ} for Cosmic Chronometers.

We use the BAO data mentioned in Section \ref{sec:observations} (labeled as BBAO); 
for Supernovae, we use the Joint Light-Curve Analysis compressed data denoted as SN, 
the Planck data (denoted as Planck) for CMB, and the Cosmic Chronometers data
(HD).

%%%================================================================%%%
\subsection{Models of the Universe}  \label{models}
%%%================================================================%%%

%%%....................................................................................%%%
\begin{table*}[t]
	\begin{center}
		\begin{tabular}{ccccccc}
		\cline{1-7}\noalign{\smallskip}
\vspace{0.2cm}
			%\multicolumn{5}{c}{\textbf{Models.}}\\
			%\hline
\qquad Parameter \qquad & \qquad $\Lambda$CDM \qquad &\qquad o$\Lambda$CDM  \qquad
    &\qquad $\omega$CDM \qquad & \qquad o$\omega$CDM \qquad
    &\qquad $\omega_0\omega_a$CDM  \qquad &\qquad o$\omega_0\omega_a$CDM \qquad \\
\hline
\vspace{0.2cm}
$\Omega_m$ & $0.299  \pm 0.007 $  & $0.298 \pm 0.007$ & $0.303 \pm 0.009$ & $0.299 \pm 0.009$
           & $0.307 \pm  0.010$ & $0.306 \pm  0.010$  \\
\hline
\vspace{0.2cm}
$\Omega_b h^2$& $0.0224 \pm 0.0002$ & $0.0227 \pm 0.0003$ & $0.0224 \pm 0.0003$ & $0.0227 \pm 0.0003$
              & $0.0224 \pm  0.0003$ & $0.0226 \pm 0.0003$  \\
\hline
\vspace{0.2cm}
$h$ & $0.684 \pm 0.006$ & $0.679 \pm 0.007$ & $0.677 \pm 0.108$ & $0.676 \pm 0.010$ 
     & $0.674 \pm  0.011$ & $0.670 \pm 0.011$  \\
\hline
\vspace{0.2cm}
$\Omega_k$ & $\cdots$ & $-0.004 \pm 0.002$ & $\cdots$ & $-0.003 \pm 0.003$
            & $\cdots$ & $-0.006 \pm  0.003$  \\
\hline
\vspace{0.2cm}
$\omega_0$ & $\cdots$ & $\cdots$ & $-0.96 \pm 0.05$ & $-0.97 \pm 0.05$
            & $-0.91 \pm  0.10$ & $-0.83 \pm  0.11$  \\
\hline
\vspace{0.2cm}
$\omega_a$ & $\cdots$ & $\cdots$ & $\cdots$ & $\cdots$
            & $-0.17 \pm  0.41$ & $-0.52 \pm 0.51$  \\
\hline
\vspace{0.2cm}
$\chi^2_{\rm min}$ & $73.57$ & $71.59$ & $73.13$ & $71.5$ & $72.8$ & $69.7$ \\
\hline
\vspace{0.2cm} 
$| \mathcal{B}_{\Lambda{\rm CDM},i}|$ & 0 & 3.61 & 1.48 & 5.59  & 1.23 & 4.3  \\
\hline
\hline
		\end{tabular}
		\caption{\footnotesize{Cosmological parameter constraints from BAO data combined with our 
		compressed description of CMB from Planck, the JLA SN 
		and Hubble data (BBAO+Planck+SN+HD). Two-tailed distributions are
		shown along with 1$\sigma$ C.L.
		Entries for which the parameter is fixed 
		are marked with dash ( $\Omega_k=0$, $\omega= \omega_0 =-1$, $\omega_a=0$).}}
		\label{tablaowaCDM}
	\end{center}
\end{table*}
%%%....................................................................................%%%

The base example corresponds to the flat $\Lambda$CDM model, already explored in 
Section \ref{example1}, but now including the full observations presented in 
Section \ref{sec:observations} and using the SimpleMC code.
%The SimpleMC code is easily manageable and its documentation can be found 
%in \cite{simplemc}.
%
Here, in order to test the Friedmann Equation \eqref{friedmann3} with the data,
we consider as free parameters (along with flat priors):
the total matter dimensionless density parameter $\Omega_m \in [0.05,1.5]$, the 
baryon physical density $\Omega_b h^2 \in [0.02,0.025]$, and the 
dimensionless Hubble constant $h \in [0.4,1]$. 
Then, assuming the base $\Lambda$CDM model, we let the curvature of the Universe be a 
free parameter
(model o$\Lambda$CDM), with its corresponding flat prior $\Omega_k \in [-1.5,1.5]$.
Moreover, because the cosmological constant is only a particular case for the 
dark energy equation-of-state $\omega=-1$, we let $\omega$ be a free parameter
with flat priors $\omega \in [-2.0,0.0]$ and labeled it as model $\omega$CDM.
We may combine the addition of curvature and constant 
$\omega$ to define the o$\omega$CDM model.
In order to go even further and describe a dynamical dark energy,
we use the CPL parameterization for the equation of state 
with flat priors on the parameters 
$\omega_0 \in [-2.0,0.0]$ and $\omega_a \in [-2.0,2.0]$
and labeled it as $\omega_0 \omega_a$CDM model. 
Again, we can incorporate the curvature 
of the Universe to the CPL parameterization, named as o$\omega_0 \omega_a$CDM.

\begin{figure}[t]
	\centering
	\includegraphics[trim = 0mm  0mm 0mm 0mm, clip, width=7.cm, height=4.5cm]{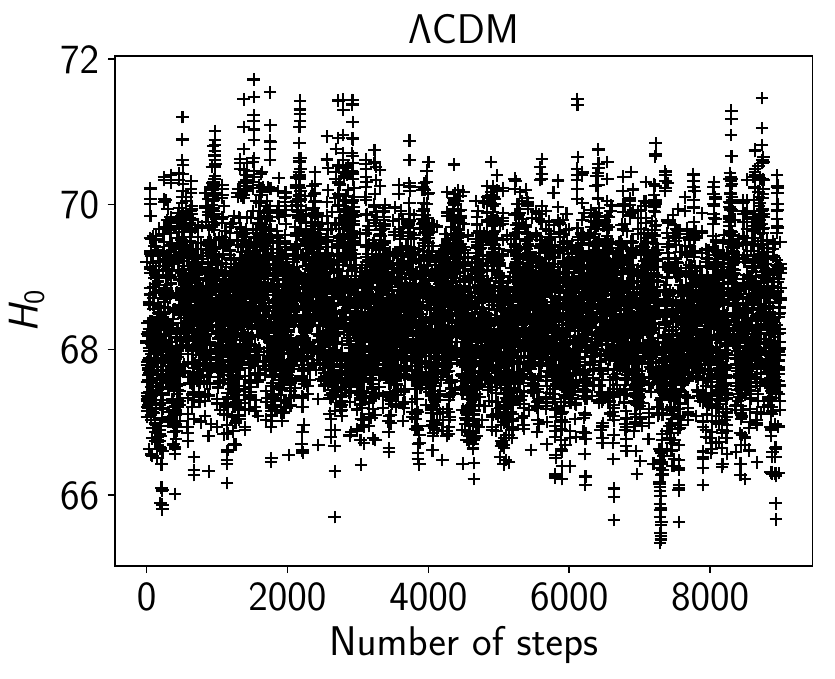} \qquad
	\includegraphics[trim = 0mm  0mm 0mm 0mm, clip, width=7.cm, height=4.5cm]{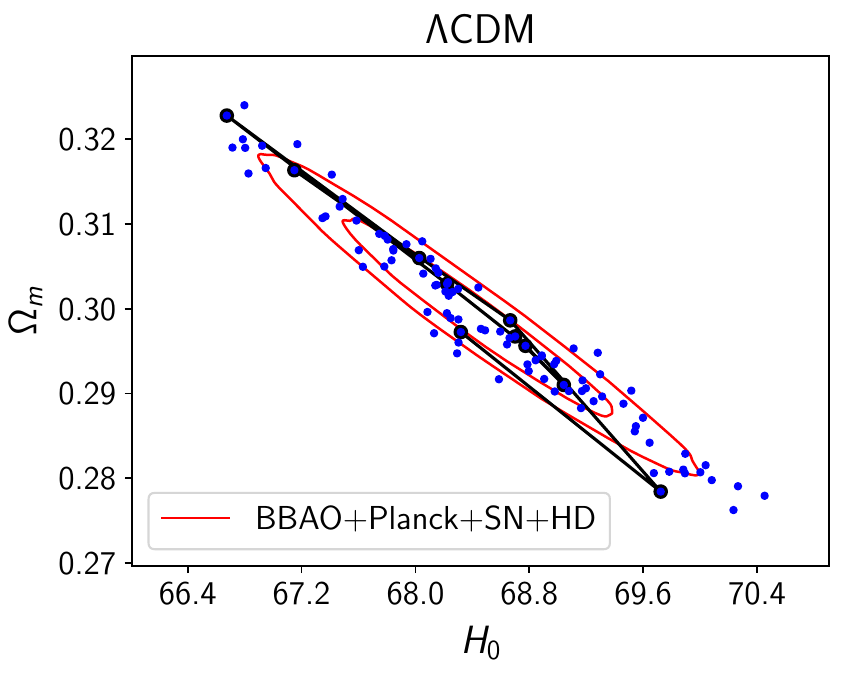}
	\caption{\footnotesize{Top panel: Markov chain for the parameter $H_0$, 
	with 9000 steps. Bottom panel: 2D posterior distribution with confidence regions 
 1 and 2 $\sigma$ for the joint parameters $H_0$ and $\Omega_m$.} }
	\label{LCDM}
\end{figure}

{By using the combined dataset BBAO+Planck+SN+HD, Table \ref{tablaowaCDM} 
shows the best fit values, along with 1$\sigma$ confidence levels.
The first important result to highlight is how the constraints have shrunk 
once new information is taken into account. That is, 
in Section~\ref{example1}, }for the $\Lambda$CDM model and 
by using only Hubble data, we had 
$h=0.677\pm 0.313$ and $\Omega_m=0.331 \pm 0.0628$.
Now, with the inclusion of BAO, Planck, and SN data, the constraints have improved considerably to $h=0.684\pm 0.006$ and $\Omega_m=0.299 \pm 0.007$.
Figure \ref{chains_lcdm} is updated with new data in order to get Figure \ref{LCDM}.
Here, the upper panel displays the chain for the parameter $H_0 = 100 \cdot h$, with 9000 steps.
In the lower panel of the same figure, we plot the 2D posterior distribution, along
with 1 and 2$\sigma$ confidence regions obtained from our estimations. 

\begin{figure}[h]
	\centering
	\includegraphics[trim = 0mm  0mm 0mm 0mm, clip, width=4.2cm, height=4.5cm]{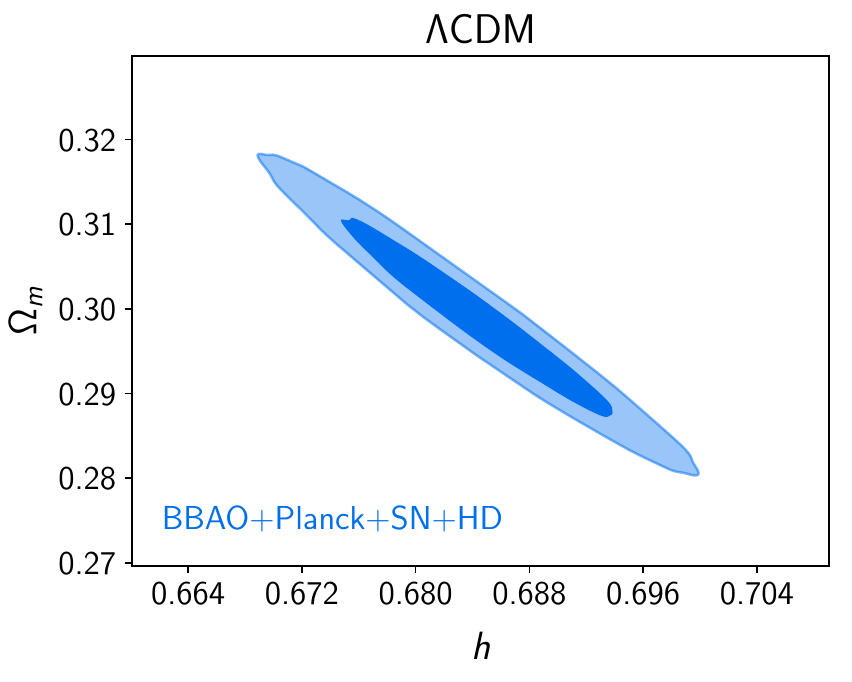}
	\includegraphics[trim = 0mm  0mm 0mm 0mm, clip, width=4.2cm, height=4.5cm]{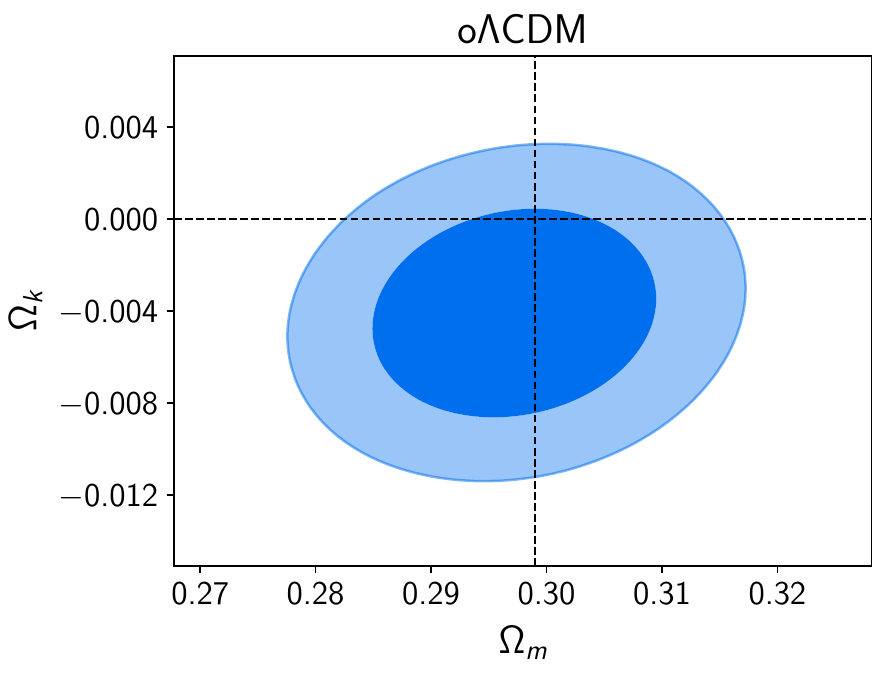} \\
	\includegraphics[trim = 0mm  0mm 0mm 0mm, clip, width=4.2cm, height=4.5cm]{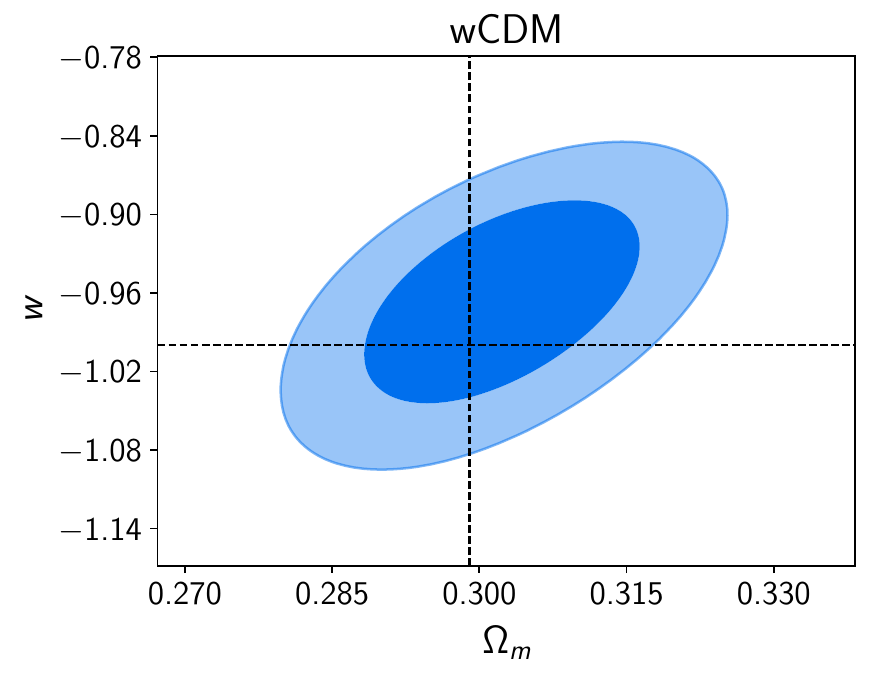}
	\includegraphics[trim = 0mm  0mm 0mm 0mm, clip, width=4.2cm, height=4.5cm]{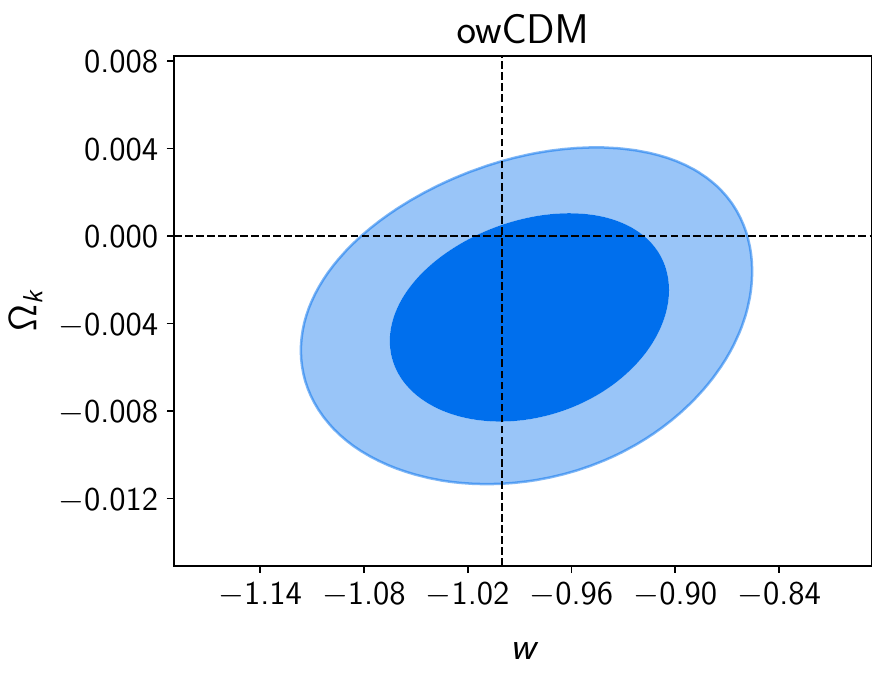}\\
	\includegraphics[trim = 0mm  0mm 0mm 0mm, clip, width=4.2cm, height=4.5cm]{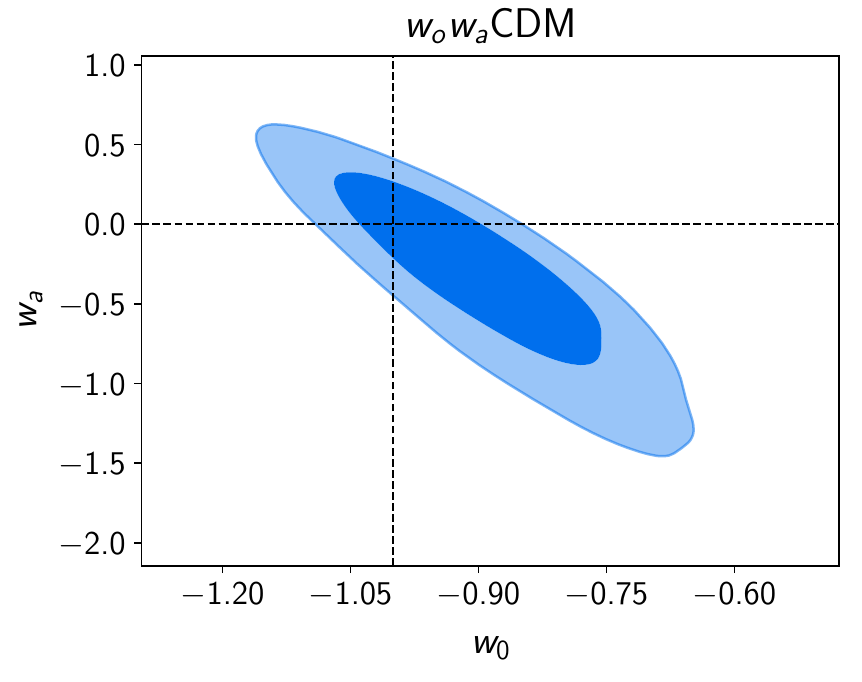}
	\includegraphics[trim = 0mm  0mm 0mm 0mm, clip, width=4.2cm, height=4.5cm]{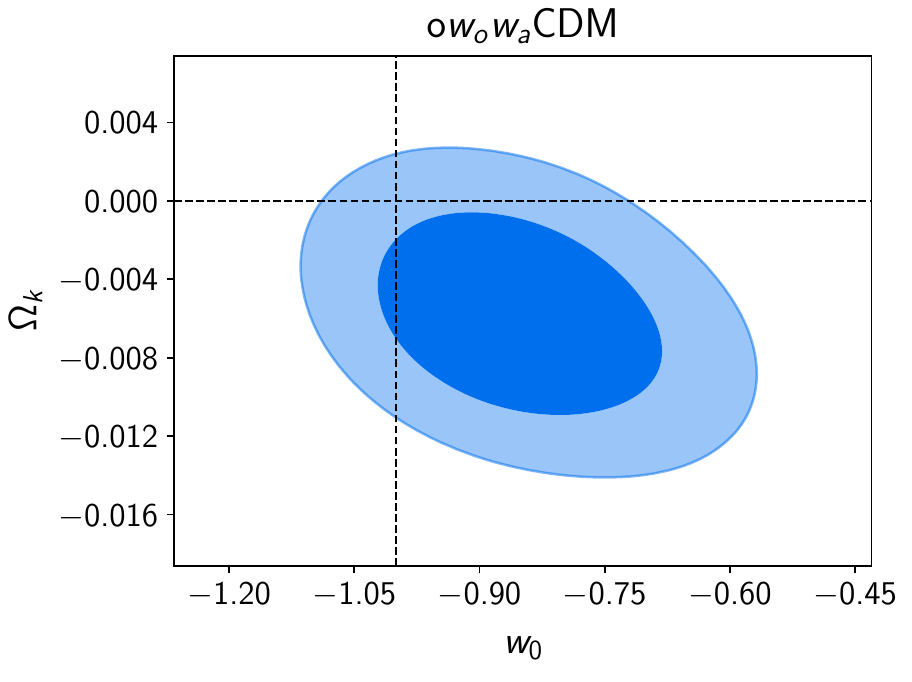} \\
	\caption{\footnotesize{2D posterior distributions 
	with 1,2-$\sigma$ confidence regions for different models. 
	o$\Lambda$CDM refers to $\Lambda$CDM with curvature,
	$\omega$CDM is a flat Universe with the dark energy equation of state as 
	a free parameter,
	o$\omega$CDM generalizes to non-zero curvature, $w_o w_a$CDM uses the CPL 
	parameterization and o$w_o w_a$CDM generalizes to non-zero curvature. 
	The dashed lines show the standard $\Lambda$CDM values. 
	}}
	\label{all_models}
\end{figure}

From Table \ref{tablaowaCDM}, we also observe the best fit of most of the new 
parameters---additional to the base $\Lambda$CDM model---remained well 
inside the 1$\sigma$ confidence level. 
However, the exception is for the o$\omega_0\omega_a$CDM model, 
where $\Omega_k$, $\omega_0$, and $\omega_a$ lay down right outside the 1$\sigma$
region. This main feature is better observed in Figure \ref{all_models}, where 
the 2D posterior distributions, along with 1 and 2$\sigma$ confidence regions,
are shown. Here, the standard $\Lambda$CDM values 
are marked with a dash line ( $\Omega_k=0$, $\omega= \omega_0 =-1$, $\omega_a=0$).
The inclusion of extra parameters improves the fit to the data, 
 observed through the minimum $\chi^2_{\rm min}$, before the last row of Table \ref{tablaowaCDM}.
However, it also carries out a penalization factor that affects directly the model selection, as seen in Reference \cite{Hee16, fourier, dynamicalDE}. 
%MDPI: Please confirm it. 
%Authors: We confirm, but we prefer the table to be closer to the text. 
% even though is not on the top of the page.

In the last row of the same table, we show the Bayes factor for the extended models using the LCDM model as reference, where the Bayesian evidences were obtained using the MCEvidence code. 
We found significant evidence in favor of $\Lambda$CDM compared to $\omega$CDM and $\omega_0\omega_a$CDM; strong evidence with respect to o$\Lambda$CDM and o$\omega_0\omega_a$CDM; and decisive evidence against o$\omega$CDM. This is in agreement with the results presented in Reference \cite{Heavens:2017hkr}, where Planck 2013 found no evidence against $\Lambda$CDM.

%%%....................................................................................%%%
%\jav{Quitar etiqueta LCDM$\_$phy...., graficar solo hasta 9000 pasos, hacer mas grandes ejes --que tenga el mismo formato que la figura de la derecha.}
%%%....................................................................................%%%
%%%....................................................................................%%%
%\jav{Panel medio deberia tener $w$ en lugar de $w_0$ }
%%%....................................................................................%%%

\section{Conclusions and Discussion}
\label{sec:Conclusions}

The fact that the number of cosmological observations have increased 
impressively over the last decade allowed to obtain a 
better description of the Universe.
However, since we still have the limitation of a unique Universe, 
a Frequentist approach may not be the best to rely on; hence, the
Bayesian statistics came into consideration.
In this work, we provide a review of the Bayesian statistics and present 
some of its applications to cosmology, mainly throughout several examples.

The Bayesian statistics rests on the rules of probability which yield
to the Bayes' theorem. Given a model or hypothesis $H$ for some data $D$, 
Bayes' theorem tells us how to determine the probability 
distribution of the set of parameters $\theta$.
 Bayes' theorem states that 
\begin{equation*}
P(\theta|D, H)=\frac{P(D|\theta,H)P(\theta|H)}{P(D|H)},
\end{equation*}
\noindent
where the {\it prior} probability $P(\theta|H)$---the state of knowledge
before acquiring the data---is upgraded through the {\it likelihood}
$P(D|\theta,H)$ when experimental data $D$ are considered.
The aim for parameter estimation is to then obtain the posterior probability
$P(\theta|D, H)$ which represents the state of
knowledge once we have taken into account the new information.

We noticed that, if  the prior probability is constant, we can identify 
the posterior probability with the likelihood $P(\theta|D, H)\propto L(D|\theta,H)$; thus, by maximizing it, we can find the most probable set of parameters 
for a model given the data. 
Moreover, if we assume a Gaussian approximation for the likelihood, 
then the chi-squared quantity is related to the Gaussian 
likelihood via $L=L_0e^{-\chi^2/2}$.
Therefore, maximizing the Gaussian likelihood is equivalent to
minimizing the chi-squared. Once the posterior distribution for a set of 
parameters, of a given model, is calculated, we show the results in the 
form of confidence regions of said parameters. 
In addition, for this particular case, in which likelihoods are Gaussian, 
Fisher's matrix can be computed according to the Hessian matrix, where 
the latter contains information about the errors 
of the parameters and their covariances. The Fisher matrix gives information 
about the accuracy of the model and allows predicting how well an experiment 
will be able to constrain the set of parameters for a given model.

On the other hand, sometimes it is difficult to know, a priori, if 
multiple datasets are consistent with each other, or whether there could be one or 
more that are likely to be erroneous.
Since there is usually this uncertainty, a way to know how 
useful a dataset is useful a dataset is may be by introducing the hyperparameter method.

may be by introducing the hyperparameter method.
These hyperparameters act as weights for every dataset in order 
to take away data that does not seem to be consistent with all of them.
Here, the key quantity to estimate the necessity of introducing 
hyperparameters to our model is given by the Bayesian evidence $P(D|H)$.

The estimation of the posterior distribution is a very computationally 
demanding process, since it requires a multidimensional exploration of the 
likelihood and prior. To carry out the exploration of the cosmological parameter space, 
we focus on Markov Chain Monte Carlo methods with the Metropolis Hastings algorithm.
A Markov process is a stochastic process (that aims to describe the temporal
evolution of some random phenomenon) where the probability distribution of the immediate
future state depends only on the present state. 
Any computational algorithm that uses random numbers is called Monte Carlo. 
Thus, the Metropolis Hastings algorithm uses a transition kernel 
to construct a sequence of points (called chain) in the parameter space 
in order to evaluate the posterior distribution of said parameter. 
The generation of the elements in a Markov chain is probabilistic by construction, 
and it depends on the algorithm we are working with. The MHA is the easiest algorithm 
used in Bayesian inference; however, to explore complex posterior distributions more
efficiently, we provide a brief description of several samplers.

Finally, we show how
Bayesian statistics is a very useful tool in Cosmology to determine, for instance, 
the combination of model parameters that best describes the Universe.
In particular, we confront the standard cosmological model ($\Lambda$CDM) to 
current observations and compare it to different models. We found the model that best fit the data, through $\chi^2_{\rm min}$, corresponds to a curved Universe with a dynamical dark energy, namely o$\omega_o \omega_a$CDM. It is important to clarify that this does not mean that the o$\omega_o \omega_a$CDM model is the final model; it merely shows that, for these particular datasets, there is an improvement in the fit when compared to $\Lambda$CDM, but further analysis and more data is necessary to give a verdict. However, adding extra parameters brings up a penalized factor seen through the Bayesian evidence; hence, the favored model becomes $\Lambda$CDM.

\begin{acknowledgements}
JAV acknowledges the support provided by FOSEC SEP-CONACYT Investigaci\'on B\'asica A1-S-21925, FORDECYT-PRONACES-CONACYT/304001/2020, and UNAM-DGAPA-PAPIIT IA104221.
LEP, LOT, and LAE were supported by CONACyT M\'exico. LEP also acknowledges sponsorship from CONACyT through grant CB-2016-282569.
\end{acknowledgements}

%%%....................................................................................%%%
%\appendix
%\section{A simple MCMC python code}
%\label{app:MCMC}

%Here we show our MCMC code written in Python. It is very simple and its purpose 
%is to help the reader to understand how to program an MCMC code from scratch. 
%However, if the reader is interested in more sophisticated algorithms PyMC3 module 
%in python may be useful \cite{pymc3}. 
%We wrote our code using the jupyter notebook which is an excellent editor when the program is not very extensive.  

%\begin{figure*}[h!]
%\includegraphics[width=0.8\textwidth]{Figures/c2.pdf}
%\end{figure*}
%\begin{figure*}[h!]
%\includegraphics[width=0.8\textwidth]{Figures/c3.pdf}
%\end{figure*}
%\begin{figure*}[h!]
%\includegraphics[width=0.8\textwidth]{Figures/c4.pdf}
%\end{figure*}

%\newpage
%\bibliographystyle{h-physrev}
%\bibliographystyle{hieeetr}
%\bibliographystyle{halpha}
%\bibliographystyle{h-elsevier}
%\bibliographystyle{kp}
\bibliographystyle{JHEP}
\bibliography{bibliography.bib}
\end{document}